\documentclass[structabstract]{aa}  
\usepackage{graphicx}
\usepackage{longtable}
\usepackage{txfonts}
\usepackage{natbib}
\def\swift{\textit {Swift}}

\def\xmmN{\textit {XMM-Newton}}

\def\lcs {light curves}

\def\sm {SM13}
\def\Sv { $S_{\rm V}$ }
\begin{document}
   \title{Long-term variability of AGN at hard X-rays}

   \author{S. Soldi\inst{1}\thanks{\email{soldi@apc.univ-paris7.fr}}
	 \and V. Beckmann\inst{2}
	 \and  W. H. Baumgartner\inst{3,4} 
	 \and G. Ponti\inst{5}
	 \and C.~R. Shrader\inst{3,6}
	 \and \\
	 P. Lubi\'nski\inst{7}
	 \and H.~A. Krimm\inst{3,6}
	 \and F. Mattana\inst{2}
	 \and J. Tueller\inst{3}
          }

   \institute{APC, Universit\'e Paris Diderot, CNRS/IN2P3, CEA/Irfu, Observatoire de Paris, Sorbonne Paris Cit\'e, 10 rue Alice Domon et L\'eonie Duquet, 
	     75205 Paris Cedex 13, France
	     \and Fran\c{c}ois Arago Centre, APC, Universit\'e Paris Diderot, CNRS/IN2P3, CEA/Irfu, Observatoire de Paris, Sorbonne Paris Cit\'e, 13 rue Watt, 75013 Paris, France
             \and NASA Goddard Space Flight Center, Greenbelt, MD 20771, USA
             \and Joint Center for Astrophysics, University of Maryland Baltimore County, Baltimore, MD 21250, USA
             \and Max-Planck-Institut f\"ur extraterrestrische Physik, Giessenbachstrasse 1, D-85748, Garching bei M\"unchen, Germany 
             \and CRESST, Universities Space Research Association and NASA GSFC, Greenbelt, MD 20771, USA
             \and Institute of Physics, University of Zielona G\'ora, Licealna 9, 65-417 Zielona G\'ora, Poland
	     }

   \date{Received 12 September 2013 / Accepted 15 November 2013}

  \abstract
   {}
  {Variability at all observed wavelengths is a distinctive property of active galactic nuclei (AGN). Hard X-rays
   provide us with a view of the innermost regions of AGN, mostly unbiased by absorption along the line of sight. Characterizing the intrinsic
   hard X-ray variability of a large AGN sample and comparing it to the results obtained at lower X-ray energies can significantly contribute
   to our understanding of the mechanisms underlying the high-energy radiation.
   }
   {\textit{Swift}/BAT provides us with the unique opportunity to follow, on time scales of days to years and with a regular sampling, 
   the 14--195~keV emission of the largest AGN sample available up to date for this kind of investigation. 
   As a continuation of an early work on the first 9 months of BAT data, we study the amplitude of the variations, and their dependence on sub-class and on 
   energy, for a sample of 110 radio quiet and radio loud AGN selected from the BAT 58-month survey. 
   }
   {About 80\% of the AGN in the sample are found to exhibit significant variability on months to years time scales, 
   radio loud sources being the most variable, and Seyfert 1.5-2 galaxies being slightly more variable than Seyfert 1, while absorbed and unabsorbed objects 
   show similar timing properties. 
   The amplitude of the variations and their energy dependence are incompatible with variability being driven at hard X-rays by changes of the absorption column density.
   In general, the variations in the 14--24 and 35--100~keV bands are well correlated, suggesting a common origin of the variability across the BAT energy band.
   However, radio quiet AGN display on average 10\% larger variations at 14--24~keV than at 35--100~keV, and a softer-when-brighter behavior for most of the Seyfert
   galaxies with detectable spectral variability on month time scale. In addition, sources with harder spectra are found to be more variable than softer ones, 
   opposite to what it is observed below 10~keV. These properties are generally consistent with a variable, in flux and shape, 
   power law continuum, pivoting at energies $\gtrsim 50 \rm \, keV$, to which a constant reflection component is superposed. 
   When the same time scales are considered, the timing properties of AGN at hard X-rays are comparable to those at lower energies, with at least some of the 
   differences possibly ascribable to components contributing differently in the two energy domains (e.g., reflection, absorption).
   }
   {}

   \keywords{Galaxies: active -- Galaxies: Seyfert -- X-rays: galaxies -- surveys
               }

   \maketitle
%

\section{Introduction}

Besides mapping the innermost regions of active galactic nuclei (AGN), the hardest X-rays $\gtrsim 15 \, \rm keV$ present
the further advantage of not being affected by absorption, provided that the hydrogen column density $N_{\rm H}$ along the line of sight 
is lower than a few $10^{23} \rm \, cm^{-2}$.
Hence, hard X-rays can be effectively used to study the intrinsic properties of AGN emission, testing the validity of unification models that in first 
approximation explain the differences between the different AGN classes as a function of the viewing angle (e.g., \citealt{beckmann12}). 
In radio quiet AGN, the hard X-ray emission is postulated to originate in a two phase medium, where the soft disk photons are Comptonized by electrons in a 
hot plasma above the disk \citep{haardt93}.
In radio loud objects, an additional contribution from inverse Compton scattering from relativistic electrons in the jet as well as synchrotron
radiation can dominate the X-ray output.

Moreover, AGN are known to exhibit variability at all observed frequencies, whose study can provide important information about the physics,
the structure and the dynamics of the emitting source.
In the softer X-ray domain ($\lesssim 10 \rm \, keV$), extensive variability studies have been carried out, especially in the recent years thanks to the 
numerous monitoring campaigns with \textit{RXTE} and \xmmN, covering from hour to year time scales.
Many important results have been obtained with these observations, as for instance the complex correlation of the X-rays with the emission at other wavelengths 
(e.g., \citealt{mchardy07,soldi07,bell11,mehdipour11,chatterjee11}) 
in particular in the optical-UV band (e.g., \citealt{shemmer03,uttley05,arevalo08b,breedt10}); the study of X-ray lag spectra (e.g., \citealt{arevalo08,zoghbi10,demarco13}); 
the correlations between variability properties and black hole mass (e.g., \citealt{czerny01,lu01,uttley02,papadakis04}) and accretion rate \citep{mchardy06,koerding07}. 
See \citet{mchardy10} for a review on X-ray variability and its scaling from stellar to super-massive black holes. 
On the other hand, the large majority of the hard X-ray variability studies performed in the last years focused on the spectral variability of AGN at different 
flux levels for single bright sources or for small samples of selected objects, thanks to pointed observations of the \textit{BeppoSAX} 
(e.g., \citealt{petrucci00,nicastro00,derosa07}), \textit{INTEGRAL} (e.g., \citealt{pian06,pian11,beckmann08,lubinski10,soldi11b,petrucci13}) and \textit{Suzaku} 
(e.g., \citealt{reeves07b,itoh08,terashima09,fukazawa11,reis12}) satellites.
These X-ray studies suggest different origins for the observed variability, as for example changes of the amount of absorption or of its ionisation, fluctuations in the 
seed photon flux, intrinsic modifications of the corona properties or geometry (or of the jet parameters, in case of radio loud AGN), a variable contribution
of different spectral components, with at least some of these scenarios being driven by accretion rate variations.

However, with the only exception being the \textit{CGRO}/BATSE instrument, detecting only a handful of AGN \citep{harmon04}, hard X-ray, long-term monitoring has 
not been possible until recent years, because of the observing strategy of the hard X-ray satellites and the limited size of the field of view of their instruments.
Furthermore, last generation and future hard-X-ray telescopes such as \textit{NuSTAR} will have small fields of view and relatively inflexible scheduling capabilities, 
so future studies may tend to be limited to either short time scales or in the number of observed objects. 
Since November 2004, the Burst Alert Telescope (BAT, \citealt{barthelmy05}) instrument on board the \swift\ satellite \citep{gehrels04} has been observing the sky
in the 14--195 keV energy range. Thanks to its large field of view of $\sim 1.4 \rm \, sr$ and to \swift's observing strategy, 
the BAT has been monitoring a large number of hard X-ray sources \citep{cusumano10,baumgartner13}, providing for the first time a long-term and
sufficiently sampled data set.
\swift/BAT light curves on different time scales, energy bands and extracted with different methods are provided through the
hard X-ray transient monitoring pages\footnote{\emph{http://swift.gsfc.nasa.gov/docs/swift/results/transients}} \citep{krimm13}, and for the 
BAT 58-month hard X-ray survey\footnote{\emph{http://swift.gsfc.nasa.gov/docs/swift/results/bs58mon/}} including more than 1000 objects,
about 60\% of which are AGN. The latter has been recently updated to cover 70 months of observations \citep{baumgartner13}.

A first study of the variability of hard X-ray selected AGN using BAT data has been presented by 
\citet{beckmann07b}. It included 44 AGN detected with high significance over the first 9 months of BAT observations. The variability study
in the 14--195~keV band has been performed using a maximum likelihood estimator and the structure function analysis. Among the main results of this work,
Seyfert 2/obscured objects were found to be more variable than Seyfert 1/unobscured ones, and an anti-correlation has been detected
between variability and luminosity. \\
Preliminary studies of the 5 years BAT observations but with limited numbers of AGN have been presented by \citet{soldi09}, \citet{ricci11b}
and \citet{caballero11}. In particular the latter study focused on 5 bright AGN, detecting spectral variability in three of them, consistent
with the electron plasma temperature decreasing with increasing flux, within the Comptonization scenario.

\citet[hereafter \sm]{shimizu13} have calculated for the first time the power density spectra (PDS) 
of 30 AGN at hard X-rays (all these objects are also included in our study), using BAT 14--150 keV data covering time scales from one week to a few years 
($\nu_{\rm PDS} \sim 10^{-8}-10^{-6} \rm \, Hz$). 
All but one PDS were found to be well fitted by a simple power law with slope $\alpha \sim -0.8$, similarly to the PDS measured in the 2--10~keV domain at time 
frequencies lower than a break frequency $1/T_{\rm B}$. 
In fact, the PDS of AGN in the 2--10 keV domain has a characteristic power law shape with index $\alpha_{\rm PDS} = -1$ for long time scales, 
steepening to $\alpha_{\rm PDS} = -2$ for times scales shorter than a source-specific break time $T_{\rm B}$ (\citealt{uttley05,gonzalez12}), 
which is found to scale with black hole mass and accretion rate following the relation described by \citet{mchardy06}. 
The lack of correlations between variability and luminosity, and variability and black hole mass in the sample presented by \citet{shimizu13} has, 
therefore, been ascribed to the long time scales probed by this study. 
No decisive evidence was found to determine whether or not AGN long-term variability is energy dependent.

We present here our variability analysis of the largest AGN sample with a sufficient hard X-ray monitoring to perform this kind of study. 
Our analysis provides an orthogonal approach to that presented by \sm, as we consider the frequency-independent variations 
of the full long-term light curves, and therefore we are able to extend our study to a much larger AGN sample, while for the PDS analysis a high 
signal-to-noise ratio is required, thus limiting the use of this technique to the 30 objects already presented by \sm.
Furthermore, we are also able to investigate the dependence of variability on energy in the hard X-ray band.  

A description of the AGN light curves in the BAT 58-month survey is presented in Sect.~\ref{section:lc}, together with the estimate of the systematic uncertainties.
In Sect.~\ref{section:var} the maximum likelihood estimator for variability is introduced and the selected sample of 110 AGN is described.
The correlation analysis between variability and the AGN fundamental parameters is detailed in Sect.~\ref{section:corr},
while Sect.~\ref{section:spe} presents the study of spectral variability within the BAT energy range.
We discuss our results (Sect.~\ref{section:discussion}) with particular emphasis on the comparison between hard X-ray and soft X-ray variability, since this can provide 
important insights on the mechanisms responsible for the observed emission across the high-energy spectrum. A summary of our conclusions is given in
Sect.~\ref{section:conclusions}.

Considering the importance of the analysed time scales in the results presented here, we will generally refer to long time scales (and low time frequencies) for 
variability studies considered to sample the PDS above the break $T_{\rm B}$, and to short time scales (and high time frequencies) below the break.
In the following we use a $\Lambda$CDM cosmology with $\Omega_{\rm M} = 0.3$, $\Omega_{\rm \Lambda} = 0.7$ and $H_0 = 73 \rm \, km \, s^{-1} \, Mpc^{-1}$.


\section{\swift/BAT light curves analysis}\label{section:lc}
Among the 1092 sources detected in the BAT 58-month survey, 625 are AGN.
We exclude from this list the 6 pairs of AGN whose sky positions are too close to safely avoid contamination of the light curves, i.e., objects with BAT coordinates
within 2 arcmin from each other.
For these 613 AGN we retrieved the publicly available light curves covering the time from the beginning of the mission 
up to May 2010, therefore including up to 66 months of data\footnote{We note that even though 58 months of BAT data were used 
for source detection, the light curves used in this work extend beyond this data set, i.e. up to 66 months.}.
The light curves have been extracted from the snapshot (i.e. a single \swift\ pointing lasting about 20 minutes) images corrected for off-axis effects.
The images were created in 8 energy bands, i.e. 14--20, 20--24, 24--35, 35--50, 50--75, 75--100, 100--150, 150--195~keV.
In order to convert count rates in the total 14--195~keV band to fluxes, we assume a Crab-like spectrum ($\Gamma = 2.15$; \citealt{tueller10}) with a flux of 
$F_{\rm Crab,14-195 \, keV} = 2.44 \times 10^{-8} \rm \, erg \, cm^{-2} \, s^{-1}$ corresponding to a count rate of 
$x_{\rm Crab, 14-195 \, keV} = 0.0418 \rm \, counts \, s^{-1} \, detector^{-1}$ (average from the Crab snapshot light curve).
The complete description of the BAT data analysis and light curve extraction is provided in \citet{tueller10} and \citet{baumgartner13}.

The reference light curves used for our analysis are obtained from the combined 14--195 keV light curves first rebinned to 30 days and then filtered to exclude points 
with exposure time shorter than one day, and data points with error bars larger than $\Delta F_{\rm cut, 30 days} = 1.7 \rm \, mCrab$. 
To derive this value, the histogram of the flux uncertainties for all light curves was built in logarithmic space, and the histogram peak and minimum identified. 
The value $\Delta F_{\rm cut, 30 days} = 2 \times \Delta F_{\rm peak} - \Delta F_{\rm min}$ (i.e., as far above the peak of the distribution 
as the minimum is below) is considered to mark the beginning of the high-value tail of the distribution of flux uncertainties 
and, therefore, is used as the cut-off value, $\Delta F_{\rm cut, 30 days}$.

\subsection{Light curve systematic errors}\label{section:sys}

Residual systematic uncertainties can affect the light curves and need to be taken into account before a variability analysis can be performed.
Among them, the most important contributions come from a component $\sigma_{\rm sys,A} = sys_{\rm A} \times F$ which is proportional to the source flux, a component 
$\sigma_{\rm sys,B} = sys_{\rm B} \times \sigma_{\rm stat}$ that can be derived from empty sky positions, and a term 
$\sigma_{\rm sys,C} = sys_{\rm C} \times \sigma_{\rm stat}$ due to summing up energy and time bins that are somehow correlated.
$\sigma_{\rm sys,A}$ is estimated using the Crab light curve under the assumption that it is constant. Fitting the Crab orbital light curve with a constant, 
a $\sigma_{\rm sys,A} = 0.06 \times F$ systematic error needs to be added in order to obtain a reduced $\chi^2$ equal to 1. A further source of uncertainty is 
expected following the finding of long-term variability in the Crab light curve (up to $\sim$7\% on year time scale; \citealt{wilson11}), 
therefore, the value of the $sys_{\rm A}$ term is likely to be overestimated here. Due to the difficulty of disentangling the systematic component from the intrinsic 
Crab variability, we prefer to choose a conservative approach and to adopt the $sys_{\rm A} = $ 0.06 value. In any case, even when reducing this term to 0.01, 
the amplitude of the variations increases on average by only 2--3\%, largely within the estimated uncertainties on the variability estimator (see Sect.~\ref{section:Sv}).

The component $\sigma_{\rm sys,B}$ is estimated by constructing orbital light curves from blank sky positions and building the histogram of all the resulting significances.
By selecting sky positions with no sources, a Gaussian distribution centered on zero for the significances is expected. Any significant deviation from a zero mean
value or from a width equal to 1 would point towards systematic uncertainties. Indeed the statistical errors need to be increased by 
$\sigma_{\rm sys,B} = 0.08 \times \sigma_{\rm stat}$ in order to obtain a Gaussian fit to the histogram of significances with width equal to 1.

Finally, from the comparison between the 14--195~keV monthly light curves extracted from the 1-month integrated mosaic images and the monthly light curves
obtained by rebinning the orbital light curves and summing up 8 energy bands over the full 14--195~keV range, 
we estimate that an additional statistical error contribution, $\sigma_{\rm sys,C} = 0.08 \times \sigma_{\rm stat}$, is needed to take into account the effects of correlated 
energy and time bins. 
The final statistical plus systematic error is given by $\sigma_{\rm tot} = \sqrt{(sys_{\rm A} \times F)^2 + [\sigma_{\rm stat} (1 + sys_{\rm B} + sys_{\rm C})]^2}$.

\begin{figure}
   \centering
   \includegraphics[width=9.3cm]{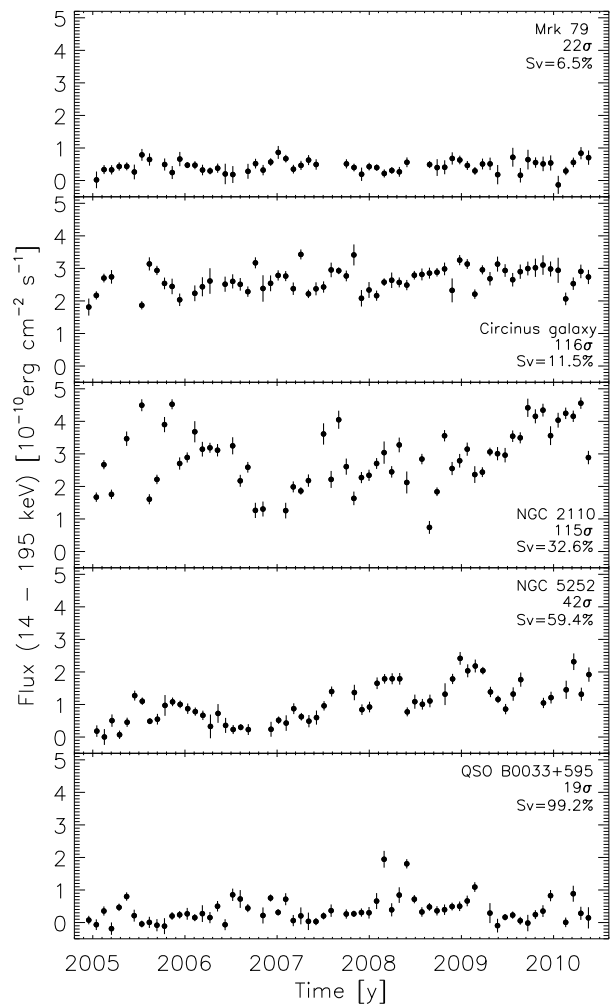}
      \caption{Examples of 30-d binned, 14--195 keV BAT light curves for objects presenting low to high variability (from top to bottom) and with different signal-to-noise ratios.
 	      The estimator \Sv\ represents the intrinsic variability of a light curve, renormalized to the average flux and in 
	      percentage units (see Sect.~\ref{section:Sv}).}
	 \label{Fig:example_lc}
   \end{figure}

In Fig.~\ref{Fig:example_lc} five example light curves are presented with different signal-to-noise ratios and variability.

   \begin{figure}
   \centering
   \includegraphics[width=5.5cm,angle=0]{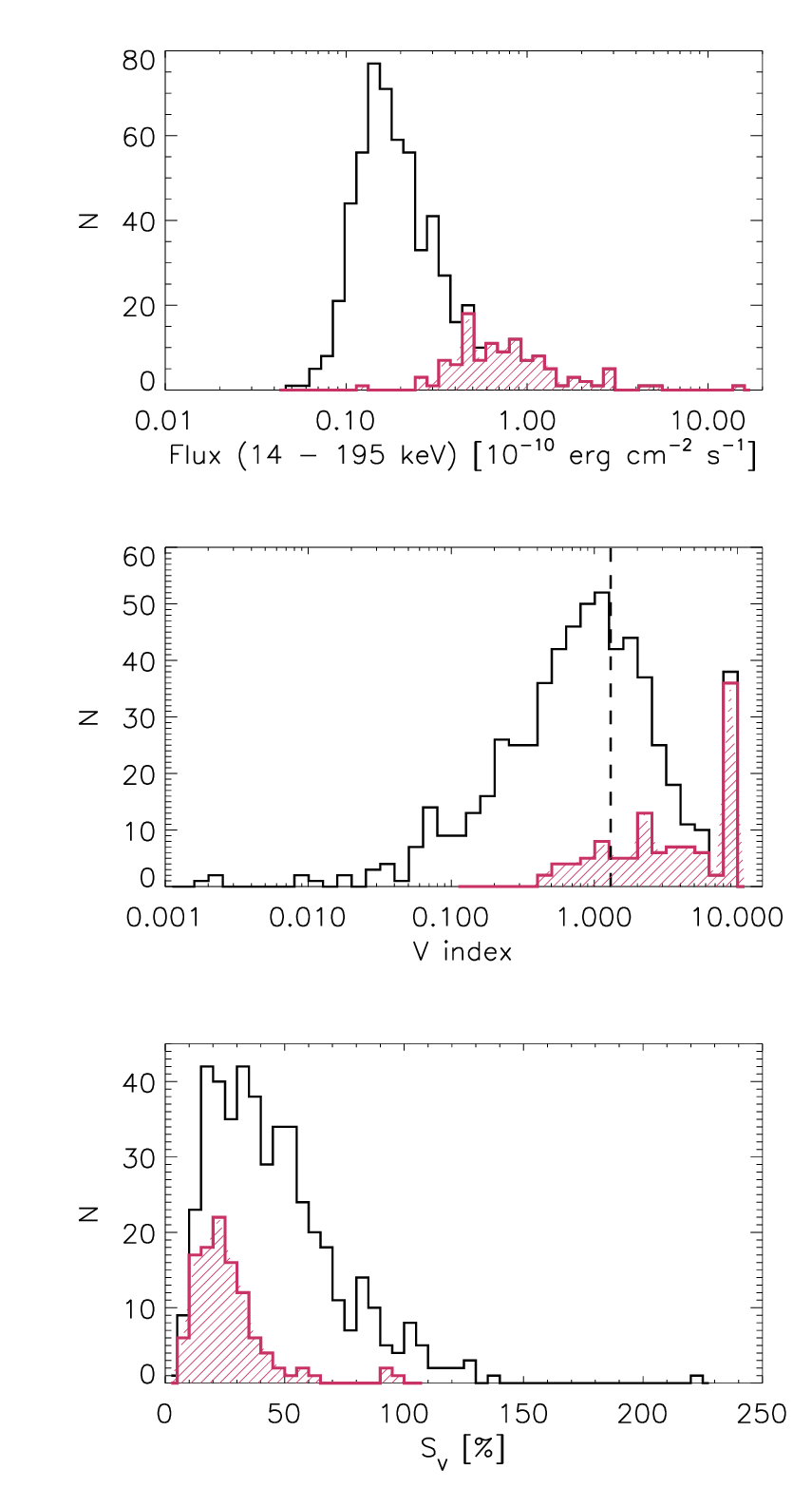}
      \caption{Histograms of the time-averaged 14--195 keV flux (top), of the variability index $V$ from the $\chi^2$ test (middle; the vertical dashed line indicates $V=1.3$), 
      		and of the variability amplitude estimator \Sv (bottom) for the total sample of 613 AGN (black) and for the selected sample of 110 AGN (red hatched). 
		We assigned $V = 10$ to those objects with $P_{\chi^2} < 10^{-6}$. \Sv could be computed for 464 of the BAT detected AGN.
	      }
	 \label{Fig:parameter_space}
   \end{figure}
%

\section{Hard X-ray variability properties}\label{section:var}
As an initial estimate of the variability of our AGN sample, we fit the 30-day binned light curves of the 613 AGN with a constant function, and
apply a $\chi^2$ test. The variability index $V$ is defined as $V = \rm -\log(1-\it P_{\chi^2})$, where $P_{\chi^2}$ is the null hypothesis probability
to obtain such a $\chi^2$ if the source were constant. For 36\% of the light curves in the sample the fit to a constant intensity source results in $V > 1.3$ 
(i.e., $P_{\chi^2} \le 5\%$; \citealt{paolillo04,lanzuisi13}), indicating that these objects exhibit significant variability (Fig.~\ref{Fig:parameter_space}). 
In particular, 34\% of the Seyfert galaxies are found to be variable against the 44\% of the radio loud population.
Among Seyfert galaxies, about 41\% of type-2 objects show variability, contrary to only 29\% of the type-1 objects.

\subsection{Amplitude of the variations}\label{section:Sv}
Several methods can be used to quantify the amplitude of the variations in a light curve, e.g., excess variance or fractional variability amplitude
\citep{vaughan03,ponti04,soldi07}. 
\citet{almaini00} proposed a numerical approach arguing that this is a more appropriate method in presence of non uniform measurement uncertainties, while recently
\citet{allevato13} showed that it provides results equivalent to the normalized excess variance.
The Almaini method is based on a maximum-likelihood estimate of the $\sigma_{\rm Q}$ parameter, 
representing the variability of the light curve and which, in the case of constant measurement uncertainties ($\sigma_i = \rm constant$), reduces to the excess variance 
($\sigma_{\rm Q} = \sigma_{\rm XS}$). The $\sigma_{\rm Q}$ parameter is defined so as to satisfy the following equation (see \citealt{almaini00} and \citealt{beckmann07b} 
for more details):
\begin{equation}
\sum^{N}_{i=1} \frac{[(x_{\rm i}-\bar{x})^2 - (\sigma_{\rm i}^2+\sigma_{\rm Q}^2)]}{(\sigma_{\rm i}^2+\sigma_{\rm Q}^2)^2} = 0
\end{equation}
where $\bar{x}$ is the mean value of the light curve $x_{\rm i}$ with measurement uncertainties $\sigma_{\rm i}$.
The maximum-likelihood estimator has been successfully used also in, e.g., \citet{mateos07} and in \citet{chitnis09}.
For continuity and comparison with \citet{beckmann07b}, we choose to apply here the Almaini method.
Unlike the original prescription, we use the weighted rather than the arithmetic mean to compute $\sigma_{\rm Q}$, because it provides in general more stable results, 
being less sensitive to the presence of outlier points. Yet, we verified that equivalent results are obtained in the two cases when analysing the final AGN sample 
presented here. 

In order to compare the variability of different objects, we renormalize $\sigma_{\rm Q}$ to the average flux $\langle F \rangle$ of the source, obtaining
the $S_{\rm V} = \sigma_{\rm Q} / \langle F \rangle \times 100\%$ variability estimator (Fig.~\ref{Fig:parameter_space}). Therefore, \Sv\ measures the amplitude of the 
intrinsic variability of a light curve, corrected by measurement uncertainties, renormalized to the average flux and in percentage units.
Differently from \citet{beckmann07b}, the systematic uncertainties deduced from the Crab and blank sky positions have already been included in the orbital \lcs\
and therefore, no further correction to \Sv\ is required.

In computing \Sv the uncertainty $\sigma_{\rm meas}$ is determined with a \textit{bootstrap} technique \citep{simpson86}. For each object, 100,000 \lcs\ are randomly 
drawn from the original one, with the same total number of points and allowing the same data point to be drawn more than once. The variability
estimator \Sv is then computed for each simulated light curve and its frequency distribution built. 
The 15.9th and 84.1th percentiles of this distribution are taken as an estimate of the 1$\sigma$ confidence intervals (see also \citealt{vaughan03}).
Due to the stochastic nature of variability, a further source of uncertainty $\sigma_{\rm sampl}$ is introduced by the uneven sampling of the light curves.
This contribution can be understood as the dispersion of \Sv values measured on the same light curve on which different samplings are applied.
We estimate this contribution with the bootstrap method but this time using the initial, unfiltered light curves (i.e., all having regular 30-day sampling). For each object, 
10,000 \lcs\ with $N$ points are randomly drawn from the original one, where $N$ is the number of points of the filtered light curve, and where every data point cannot be 
drawn more than once.
The 15.9th and 84.1th percentiles of the \Sv frequency distribution are taken as an estimate of the 1$\sigma$ confidence intervals.
The final uncertainty on \Sv is obtained by combining in quadrature $\sigma_{\rm meas}$ and $\sigma_{\rm sampl}$.

\subsection{Sample selection}
The final sample of 110 AGN is selected based on the following two criteria: 1) the average value of the signal-to-noise ratio of the points in the light curve is larger than 2;
2) \Sv can be measured for the given 30-day binned, filtered light curve, i.e., the uncertainties are smaller 
than the measurable intrinsic variability. The first criterion is essentially (but not exactly) equivalent to using a flux threshold of 
$F(14-195 \rm \, keV) > 3.8 \times 10^{-11} \rm \, erg \, cm^{-2} \, s^{-1}$ or a signal-to-noise ratio over the full light curve larger than 15, 
and it selects 115 sources (Fig.~\ref{Fig:parameter_space}). The second criterion excludes only 5 additional sources.
The filtering applied to the light curves (see Sect.~\ref{section:lc}) excludes no more than 22\% ($\sim$10\% on average) of the data for the objects in the selected 
sample.\footnote{The values of \Sv computed on the filtered and unfiltered light curves are well compatible with an average scatter of $\leq 2\%$ for the selected sample, 
well within the estimated uncertainties on $S_{\rm V}$. On the other hand, for the BAT AGN excluded from the analysis the scatter is on average 4 times larger, 
further justifying our choice to limit the analysis to a sample of light curves for which the variability measurement is not so strongly influenced by single data points.}
The resulting sample is composed of 88 Seyfert galaxies (36 Seyfert 1, 17 of intermediate type, 32 Seyfert 2, and 3 Narrow Line Seyfert~1), 9 radio galaxies and 13 blazars. 
These sources cover a range of 14--195~keV fluxes between $F = 10^{-11}$ and $10^{-9} \, \rm erg \, cm^{-2} \, s^{-1}$  (Fig.~\ref{Fig:parameter_space})
and redshifts up to $z = 2.5$ (see Fig.~1 in \citealt{soldi12} for the redshift distribution of the sample).
The non-blazar objects belong to the local AGN population with redshifts up to $z = 0.1$ and an average value of $\langle z \rangle = 0.024$ (Table~\ref{table:average}). 

In Table~\ref{table:Sv} the values of \Sv are reported for the 110 AGN selected for the variability amplitude analysis.
The blazar population shows a larger average variability ($\langle S_{\rm V} \rangle_{\rm bla} = 33 \pm 2 \%$)
compared to the radio quiet objects ($\langle S_{\rm V} \rangle_{\rm Sey} = 19.3 \pm 0.5 \%$). 
A Kolmogorov-Smirnov test provides a probability $P_{\rm KS} \geq 98 \%$ that the blazar and Seyfert samples are not drawn from the same parent population.
Yet, extreme cases of high variability are present in both the blazar and Seyfert classes. An average variability $S_{\rm V} > 90\%$ is in fact detected in the 
two gamma-ray bright blazars QSO~B0033$+$595 and Mrk~421 and in the Seyfert~2 galaxy 
2MASX~J04440903$+$2813003.

The radio galaxies present an intermediate behavior between blazars and Seyferts, with $\langle S_{\rm V} \rangle_{\rm RG} = 24.0 \pm 1.4 \%$.
Within the Seyfert class, intermediate and type 2 objects ($\langle S_{\rm V} \rangle_{\rm Sey1.5+2} = 20.8 \pm 0.6 \%$) are found to be only slightly more variable than type 1 
($\langle S_{\rm V} \rangle_{\rm Sey1} = 16.5 \pm 0.9 \%$), 
and a KS-test probability $P_{\rm KS} \geq 94 \%$ indicates that the Seyfert 1 and Seyfert 1.5--2 samples could be drawn from the same parent distribution.
Furthermore, no difference is found in the average variability of absorbed and unabsorbed Seyfert galaxies 
($N_{\rm H} = 10^{22} \, \rm cm^{-2}$ being the dividing line), with $\langle S_{\rm V} \rangle_{\rm abs} = 19.7 \pm 0.6 \%$ and 
$\langle S_{\rm V} \rangle_{\rm unabs} = 19 \pm 1 \%$.

The average properties of the different AGN classes are summarized in Table~\ref{table:average}.

\subsection{Black hole mass, hydrogen column density and Eddington ratio}\label{sect:bl_Nh_Edd}

In order to test the dependence of variability on different AGN parameters, we collected from the literature the values of black hole masses $M_{\rm BH}$ and hydrogen column 
density $N_{\rm H}$ for our sample. 
The masses have been calculated with different methods, and when not available in the literature, the uncertainties on the values were estimated from the typical accuracy 
provided by the method following \citet{beckmann09}. Out of the 110 objects, 92 have mass estimates (only an upper limit for MR~2251$-$178; Table~\ref{table:Sv}), 
86 have $N_{\rm H}$ values, and for 23 only $N_{\rm H}$ upper limits could be found.

For 63 objects (56 Seyferts, 4 radio galaxies and 3 blazars) we collected the bolometric luminosities from the literature, selecting those estimated from the fitting 
of the spectral energy distributions (\citealt{woo02,vasudevan07,vasudevan09,vasudevan09b,vasudevan10}; Table~\ref{table:Sv}). 
In order to estimate the bolometric luminosity from the X-ray one, constant corrections \citep{winter09,beckmann09}, and corrections depending on the luminosity 
\citep{marconi04}, on the Eddington ratio \citep{vasudevan07}, on the spectral shape \citep{fabian09}, or on the intrinsic absorption \citep{vasudevan10} have been proposed, 
but all fail to reproduce the large scatter of bolometric versus X-ray luminosity.
However, in order to extend the sample for which we are able to compute the Eddington ratio, we also considered two of those bolometric corrections. 
The first one follows \citet{beckmann09}, i.e., assumes that the bolometric luminosity $L_{\rm bol}$ is 2 times larger than the 
1--200~keV luminosity, where $L_{\rm 1-200 \rm \, keV} = 1.8 \times L_{\rm 14-195 \rm \, keV}$ for a simple power law spectrum with photon index $\Gamma = 1.9$.
The second one is based on the relation provided by \citet{marconi04} to compute $L_{\rm bol}$ using a luminosity-dependent bolometric correction. 
Since these corrections allow to have estimates of $L_{\rm bol}$ for additional 29 objects (19 Seyfert) and they trace sufficiently well the bolometric luminosities 
computed from the SED fitting for our sample, we consider $L_{\rm bol, \, SED}$ throughout the paper but compare the results to those obtained with X-ray based 
$L_{\rm bol}$ estimates on a larger sample.
The Eddington ratio is then computed as $\lambda_{\rm Edd} = L_{\rm bol} / L_{\rm Edd}$, where $ L_{\rm Edd} = 1.26 \times 10^{38} M_{\rm BH} / M_{\sun} \rm \, erg \, s^{-1}$. 


   \begin{figure}
   \centering
   \includegraphics[width=6.0cm,angle=90]{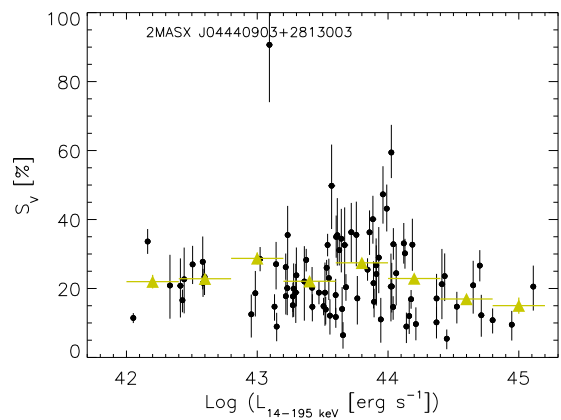}
      \caption{Variability estimator \Sv versus luminosity in the 14--195~keV band for the Seyfert galaxies in our sample.
      	       Yellow triangles represent the average of the variability estimator in different luminosity bins following \citet{allevato13}.
	      }
	 \label{Fig:Sv-lum}
   \end{figure}
   \begin{figure}
   \centering
   \includegraphics[width=6.0cm,angle=90]{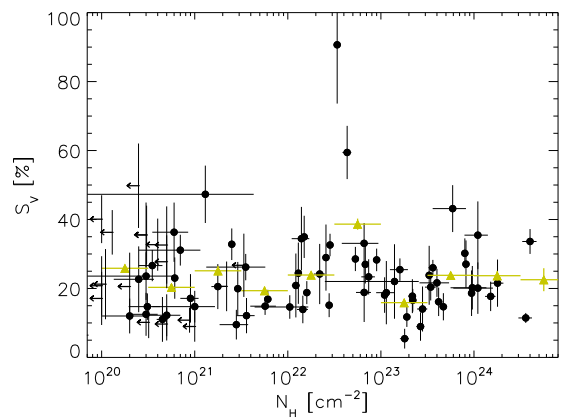}
      \caption{Variability estimator \Sv versus intrinsic absorption for the Seyfert galaxy sample. 
      		Upper limits on $N_{\rm H}$ are indicated with arrows. Yellow triangles represent the average of the variability estimator in different 
		absorption bins.
	      }
	 \label{Fig:Sv_Nh}
   \end{figure}
   \begin{figure*}
   \centering
   \hspace{-1.0cm}
   \includegraphics[width=19.0cm,angle=0]{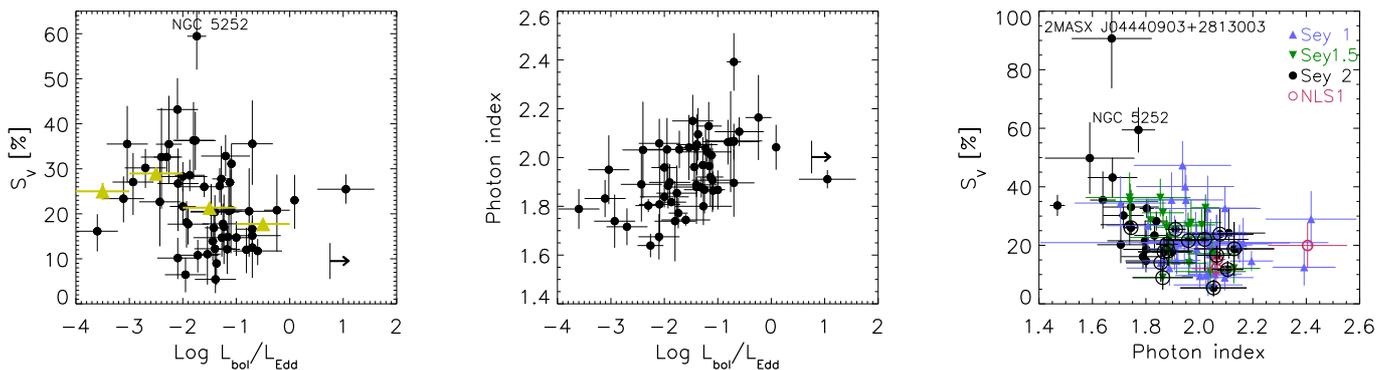}
      \caption{\textit{Left:} Variability estimator \Sv versus Eddington ratio for the Seyfert sample. Yellow triangles represent the average of the variability 
      		estimator in different Eddington ratio bins.
      		\textit{Middle:} Photon index versus Eddington ratio for the Seyfert sample.
      		\textit{Right:} Variability estimator \Sv versus photon index for all the Seyfert galaxies in our sample, separated into the different types. 
		 Empty black circles indicate those sources with $10^{23} < N_{\rm H} < 4 \times 10^{23} \rm \, cm^{-2}$.	       
	     }
	\label{Fig:Gamma_Sv}
   \end{figure*}

\section{Correlation analysis}\label{section:corr}

In order to test for the dependence of the hard X-ray variability on the properties of the selected objects, we correlate the variability
estimator \Sv with the X-ray luminosity $L_{\rm 14-195 \, keV}$, black hole mass $M_{\rm BH}$, Eddington ratio $\lambda_{\rm Edd}$, intrinsic hydrogen column density 
$N_{\rm H}$ and spectral shape $\Gamma$. 
We exclude from the correlation analysis the radio loud sources in order to have a uniform, local Seyfert sample for which it is not necessary 
to correct variability and luminosity for redshift and beaming effects, and for which the hard X-ray emission is dominated by accretion processes, with no 
jet contribution. In addition, the results for the Seyfert sample will be easily comparable to previous studies also focusing on radio quiet objects.

We do not find any significant correlation of \Sv with luminosity (Fig.~\ref{Fig:Sv-lum}) nor with absorption (Fig.~\ref{Fig:Sv_Nh}).
Variability appears to be marginally anti-correlated with Eddington ratio when using $L_{\rm bol, \, SED}$ to compute $\lambda_{\rm Edd}$,
with a Spearman ranking correlation coefficient of $R_{\rm corr} = -0.3$ corresponding to a probability for chance occurrence of $P_{\rm corr} = 0.03$ 
(Fig.~\ref{Fig:Gamma_Sv}, left panel). However, the correlation disappears when using a larger sample and the X-ray based bolometric corrections to estimate $\lambda_{\rm Edd}$. 
When fitting the \Sv versus Eddington ratio relation with a power law model, we find $S_{\rm V} \propto \lambda_{\rm Edd}^{-0.044 \pm 0.014}$, which steepens to
$S_{\rm V} \propto \lambda_{\rm Edd}^{-0.09 \pm 0.02}$ when excluding the 3 sources at super-Eddington rates.

\begin{figure}
   \centering
   \includegraphics[width=6.1cm,angle=90]{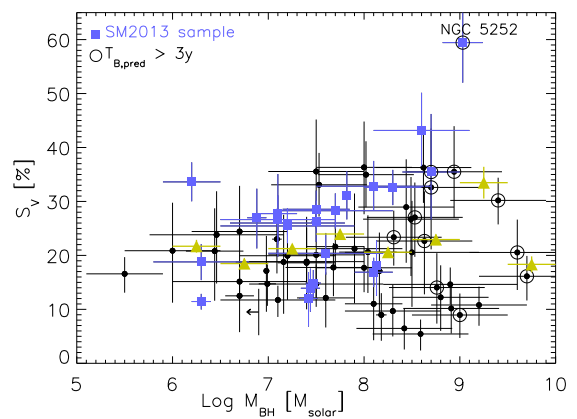}
      \caption{Variability estimator \Sv versus black hole mass for the Seyfert galaxies in the sample.
      Empty circles highlight AGN with predicted break time larger than 3 years, while blue empty squares point out the Seyferts from the \sm\ sample.
      The arrow indicates the $M_{\rm BH}$ upper limit for MR~2251$-$178.
      Yellow triangles represent the average of the variability estimator in different black hole mass bins.
	      }
	 \label{Fig:Sv_Mbh}
   \end{figure}

Unlike what is observed in the 2--10 keV band, we do not detect any anti-correlation between the 14--195 keV variability and the black hole mass 
($R_{\rm corr} = 0.06$, $P_{\rm corr} = 0.64$; Fig.~\ref{Fig:Sv_Mbh}). This is in agreement with what was 
found by \citet{shimizu13} on a smaller sample (squares in Fig.~\ref{Fig:Sv_Mbh}). 
On the other hand, \citet{caballero11} reported a trend of decreasing variability with increasing black hole mass for the 5 brightest Seyfert galaxies 
at hard X-rays, with 20--50 keV BAT light curves rebinned to 2 days. However, when computing $F_{\rm var}$ for the 2-days binned, 14--195 keV light curves 
as presented in our work for the same 5 sources, and applying a Spearman rank correlation test (more robust than the linear/Pearson correlation 
one; e.g., \citealt{press07}), we do not find any anti-correlation between variability and black hole mass ($R_{\rm corr} = -0.1$, $P_{\rm corr} = 0.87$).

The lack of a $S_{\rm V} - M_{\rm BH}$ anti-correlation is most likely due to the fact that the BAT survey samples time scales longer than the power density 
spectrum (PDS) break, $T_{\rm B}$. In fact, the X-ray PDS of AGN has a characteristic power law shape with index $\alpha_{\rm PDS} = -1$ for long time scales 
and $\alpha_{\rm PDS} = -2$ for times scales shorter than a source-dependent break time $T_{\rm B}$ (typically of the order of hours to weeks; 
\citealt{uttley05,arevalo08,markowitz09,markowitz10,gonzalez12}), 
which is found to scale with black hole mass and bolometric luminosity following the relation described by \citet{mchardy06}. This implies a scaling of the variability amplitude 
with the same AGN properties when time scales of the order or shorter than $T_{\rm B}$ are considered.
Above 15~keV the PDS obtained with BAT data on time scales of weeks to years show for 29 out of 30 objects a single power law with index $\alpha_{\rm PDS} \sim -0.8$ (\sm), 
consistent with the 2--10~keV results and providing an upper limit of 26 days for the hard X-ray $T_{\rm B}$ of these objects.
However, if one assumed that $T_{\rm B}$ is energy independent and can be estimated using the \citet{mchardy06} relation, one would find that for 12 Seyferts in our sample 
$T_{\rm B,\, pred}$ is predicted to be larger than 3 years. This suggests the possibility that for those AGN BAT has measured the variability at frequencies above the 
PDS break, and that therefore a scaling of variability with black hole mass should be expected\footnote{For the majority of the objects in the \sm\ sample (20 out of 23) 
$T_{\rm B,\, pred}$ is of the order or lower than 30 days, consistent with no $T_{\rm B}$ detected by \sm\ in the BAT time frequency range.}.
Yet, no correlation is detected (empty circles in Fig.~\ref{Fig:Sv_Mbh}). 
Recently, based on the 0.2--10~keV PDS analysis of 104 AGN and on the detection of a break in 15 of them, \citet{gonzalez12} recomputed the 
$T_{\rm B}-M_{\rm BH}-L_{\rm bol}$ relation finding a weaker dependence on $L_{\rm bol}$ than in the \citet{mchardy06} formulation.
Therefore, considering the large uncertainties on the mass determination and bolometric luminosity estimate, 
and on the $T_{\rm B}-M_{\rm BH}-L_{\rm bol}$ relation, it is not possible to derive from the above arguments whether or not this relation is also 
valid at hard X-rays.

\section{Spectral variability}\label{section:spe}

\subsection{Variability, photon index and Eddington ratio}
A significant anti-correlation is found between variability and photon index for Seyfert galaxies, in the sense that AGN with harder spectra are more variable
($R_{\rm corr} = -0.5$, $P_{\rm corr} = 2 \times 10^{-6}$, Fig.~\ref{Fig:Gamma_Sv} right panel). 
The photon index used here is the result of a simple power law fit to the BAT spectrum. 
Even when excluding the objects with $N_{\rm H} > 3 \times 10^{-23} \rm \, cm^{-2}$ for which absorption might start to have a noticeable influence
on the hard X-ray spectrum, the anti-correlation is still significant ($P_{\rm corr} = 6 \times 10^{-4}$).

A correlation is also identified between photon index and Eddington ratio ($R_{\rm corr} = 0.56$, $P_{\rm corr} = 10^{-5}$; Fig.~\ref{Fig:Gamma_Sv} middle panel), 
similar to what is observed in the 2--10~keV band on year time scales \citep{sobolewska09}. The best fit to the data with a power law model results in 
$\Gamma \propto \lambda_{\rm Edd}^{0.022 \pm 0.002}$ ($\Gamma \propto \lambda_{\rm Edd}^{0.039 \pm 0.003}$ when considering only sources with $\lambda_{\rm Edd} < 1$). 
The correlation is significant also when using the X-ray bolometric corrections to compute $\lambda_{\rm Edd}$.
This relation has been suggested to origin from the geometry of the accretion flow, with stronger accreting objects having accretion disks whose inner radius is 
closer to the innermost circular stable orbit and therefore can cool more efficiently the hot corona, resulting in softer X-ray spectra \citep{sobolewska09}.

The marginal anti-correlation between variability and Eddington ratio (Fig.~\ref{Fig:Gamma_Sv}, left panel) could therefore be induced by the combination of the 
$\Gamma-\lambda_{\rm Edd}$ and $S_{\rm V}-\Gamma$ ones ($S_{\rm V} \propto \Gamma^{-2.5 \pm 0.3}$ for the total Seyfert sample).

\subsection{Variability in the 14--24 and 35--100~keV bands}
In order to compare the variability at different energies, we use the 30-day binned \lcs\ in the 14--24 and 35--100~keV bands (see Table~\ref{table:Sv}).  
These energy bands have been selected to provide comparable significance for a typical hard X-ray AGN spectrum with power law shape and photon index $\Gamma = 1.8$.

\begin{figure}
   \centering
   \includegraphics[width=9.0cm,angle=0]{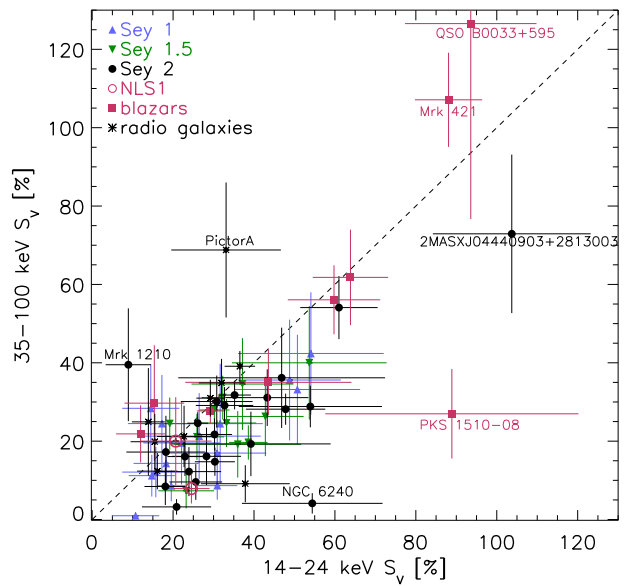}
      \caption{35--100~keV versus 14--24~keV variability estimator for the AGN in our sample. 
      The dashed line indicates where objects with the same variability in the two bands would lie.
      Even though the variations in the two bands are well correlated, for the majority of the radio quiet
      objects the variations in the lower-energy band are larger than those at higher energies. Some of the most extreme outliers are labeled.
	      }
	 \label{Fig:Sv_band}
   \end{figure}

For a sample of 68 AGN (50 Seyferts, 9 radio galaxies, 9 blazars), we are able to estimate \Sv for the two selected bands (Fig.~\ref{Fig:Sv_band}). 
In general, the variations are well correlated between the two bands ($R_{\rm corr} = 0.6$, $P_{\rm corr} = 5 \times 10^{-7}$). 
However, there is an average 10\% shift of the Seyfert population towards higher amplitude of the variations at lower energies, aside from a very few exceptions. 
A Kolmogorov-Smirnov test indicates a probability $P_{\rm KS} \geq 98 \%$ that the variations in the two energy bands are different for the 50 Seyferts in this subsample. 
This is also true when separating the sample into the 13 unabsorbed ($N_{\rm H} < 10^{22} \rm \, cm^{-2}$) and 26 absorbed objects: absorbed and unabsorbed sources show
the same average variability amplitude, with the lower energy band presenting larger variability ($\langle S_{\rm V} \rangle_{\rm unabs, 14-24 \, keV} = 26 \pm 2 \%$, 
$\langle S_{\rm V} \rangle_{\rm abs, 14-24 \, keV} = 28 \pm 1 \%$, $\langle S_{\rm V} \rangle_{\rm unabs, 35-100 \, keV} = 17 \pm 2 \%$, 
$\langle S_{\rm V} \rangle_{\rm abs, 35-100 \, keV} = 16 \pm 1 \%$). On the other hand, on average the 18 radio loud AGN do not display a significant dependence 
of variability on energy.

It is important to stress that the 14--24 and 35--100~keV bands have been selected specifically to have similar signal-to-noise ratio (SNR) for a typical AGN spectrum, 
therefore excluding the possibility that the behavior observed could be due to a bias introduced by the different uncertainties in the two bands.
In fact, on average the difference between the SNR in the low- and high-energy band is 3\% for the total sample ($-1.4\%$ for the Seyferts and 11\% for the radio loud sources), 
with about half of the sample having a larger SNR in the high-energy band.

When testing the possibility that the additional variability in the lower energy band could be ascribed to variations of the intrinsic absorption, we find that large and 
frequent variations would be needed. In fact, if we consider a simple power law spectrum ($\Gamma = 1.9$) with full-coverage absorption, variations of the order of 
$\Delta N_{\rm H} \sim 5-10 \times 10^{23} \, \rm cm^{-2}$ would be necessary to produce flux variations of $\Delta F_{14-24 \, \rm keV} \sim 8-16\%$ and 
$\Delta F_{35-100 \, \rm keV} < 0.7\%$. This would imply a 10\% difference between the low- and high-energy flux variations within two observations and correspond 
to a $\Delta F_{14-195 \, \rm keV} \sim 3-6\%$ contribution to the variability in the full band.
However, in order to measure an increase in \Sv by 10\% given our sampling baseline and rate, unrealistic large ($\langle \Delta N_{\rm H} \rangle$ at least up to 
$\sim 10^{25} \rm \, cm^{-2}$) and frequent (on average over 50\% of the light curve points) variations in the absorbing column would be required, and would imply 
even larger variations in the energy band $< 10 \, \rm keV$.
Even though variable absorbers are indeed common \citep{risaliti02c}, such extreme variations are very rarely observed (see e.g., NGC~1365, \citealt{risaliti07}).

   \begin{figure*}
   \centering
   \includegraphics[width=18.0cm,angle=0]{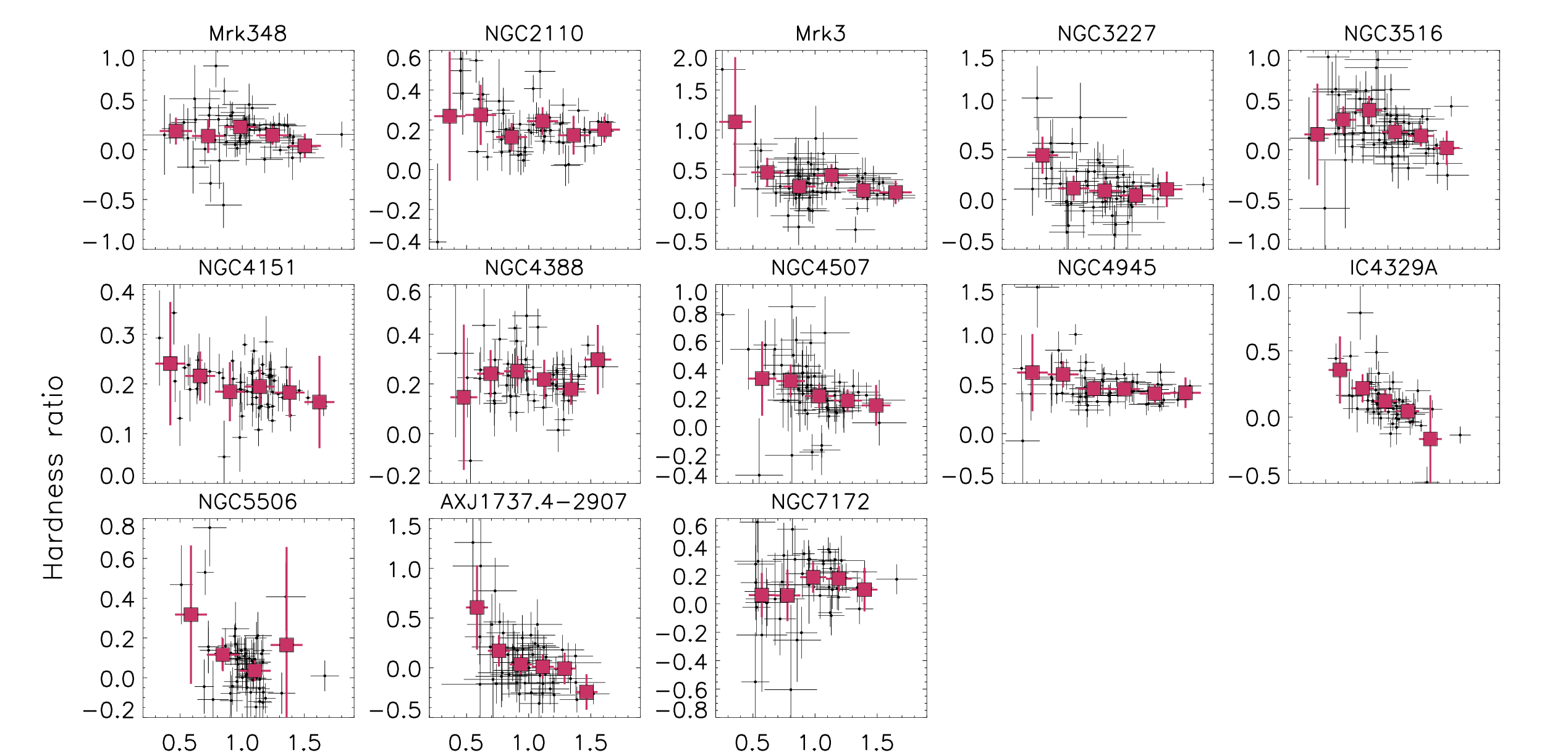}
   \includegraphics[width=18.0cm,angle=0]{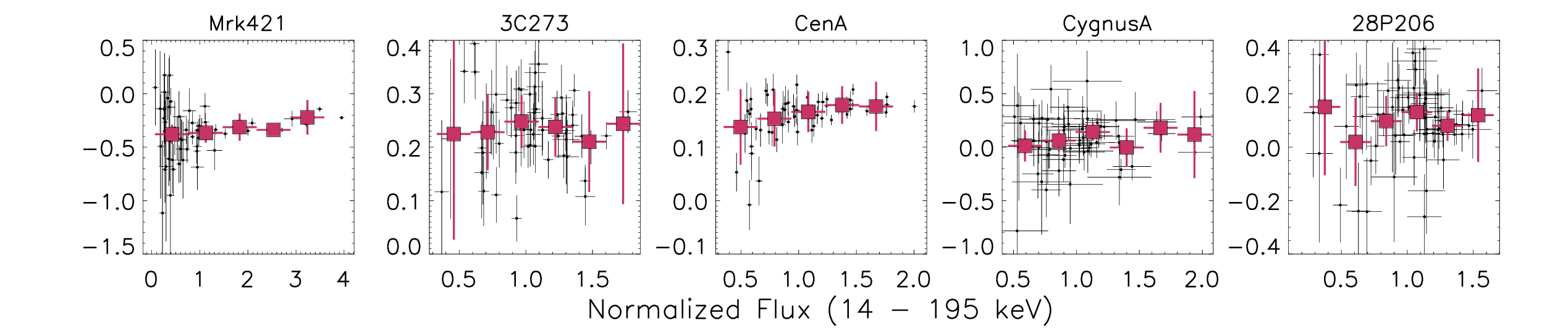}
      \caption{Hardness ratio versus 14--195~keV flux for the 18 AGN showing spectral variability (13 Seyfert galaxies on the three top rows and 5 radio loud AGN on the 
      		bottom row). The black dots are from the 1-month binned light curves, and red squares indicate the average hardness ratio in different flux bins.
	      }
	 \label{Fig:HR_flux}
   \end{figure*}

\subsection{Hardness ratio versus flux}
In order to compare the spectral variability within the single sources, for each object we compute the hardness ratio on month time scale, defined as: 
\begin{equation}
HR = \frac{F_{35-100 \rm \, keV}-F_{14-24 \rm \, keV}}{F_{35-100 \rm \, keV}+F_{14-24 \rm \, keV}} \,\,\,\,\,\,\,\,\, .
\end{equation}
The dependence of $HR$ on time is then fitted with a constant function, a $\chi^2$ test is applied, and the corresponding variability index $V$ is computed 
(as defined in Sect.~\ref{section:var}). Based on this, we select a subsample of 18 objects for which $V > 1.3$, as those presenting spectral variability. 
These are mostly the brightest objects in the sample, with a few exceptions (e.g. MCG$-$05$-$23$-$016, Circinus galaxy).
The relatively low fraction of sources showing significant spectral variability is in agreement with the results presented by \citet{mateos07} who found 
spectral variability to be less common than flux variability on long time scales in the 0.2-12~keV band.
In Fig.~\ref{Fig:HR_flux} the hardness ratios as a function of flux (normalized to the mean flux of each object) are presented. For 8 Seyferts there is a clear trend of 
spectral softening when the source brightens (Mrk~3, NGC~3516, NGC~4151, NGC~4507, NGC~4945, IC~4329A, NGC~5506, AX~J1737.4$-$2907), while for the other radio quiet 
objects no significant correlation is detected. Our results are in agreement with those reported by \citet{caballero11} for NGC~4945, NGC~2110 and IC~4329, 
while different trends are observed for NGC~4388 and NGC~4151. This is possibly due to the different energy bands used to compute $HR$ 
(20--50 and 50--100 keV in \citealt{caballero11}).

There is no correlation between flux and spectral variations for radio loud objects, with the exception of Mrk~421 and Cen~A, which exhibit a marginal spectral hardening 
with increasing flux. Cen~A has been observed to present a harder-when-brighter spectrum also during \textit{Suzaku} observations \citep{fukazawa11}. 
This behavior has been interpreted as due to the increased contribution in the brighter state of a hard tail produced in the jet.


\section{Discussion}\label{section:discussion}

\subsection{Variability of the different AGN classes}

Radio loud objects show larger amplitudes of the variations compared to Seyfert galaxies, as expected if their hard X-ray emission is dominated by the extreme processes 
taking place in a relativistic jet (e.g., \citealt{lichti08}), while hard X-ray emission from Seyfert galaxies is instead dominated by thermal Comptonization. 
This is even more evident considering that the variability of higher redshift blazars (4 out of 13 objects have $z > 0.4$) will actually increase when taking into 
account the effect of cosmological time dilation.
Among radio loud objects, the radio galaxies seem to show rather intermediate variability amplitudes between the blazars and the radio
quiet class, in agreement with the idea that the jet and the coronal emissions are probably both contributing. This makes it difficult to disentangle the relative importance
of these components in the hard X-ray spectra of radio galaxies, as for example shown in Centaurus~A \citep{beckmann11}, 3C~111 \citep{dejong12}, as well as in other 
radio galaxies \citep{grandi06}. 

Among Seyfert galaxies, there are indications for type 1.5--2 objects to be slightly more variable than type 1, as found already at hard X-rays in \citet{beckmann07b}. 
A similar behavior has been reported by \citet{saxton11} when studying long-term X-ray variability properties of a sample of more than 1000 AGN detected by 
\textit{ROSAT} and within the \xmmN\ slew survey. Following their study, the Seyfert 2 class has the highest fraction of sources with large variability in the 
0.2--2~keV range, but it is not clear if the observed variability is due to changes in the line-of-sight column density, or it is rather intrinsic to the 
central engine.
On the other hand, while in \citet{beckmann07b} there was a tentative trend of more absorbed sources being more variable, here we do not observe any correlation 
between variability and absorption column density. Therefore, the global properties of variability as a function of AGN class point to a general agreement with the 
basic formulation of AGN unification models, in which the differences are driven in first approximation only by AGN orientation and by the presence or absence of a 
relativistic jet.

Within our sample, there are 3 NLS1 galaxies, a class which is known to show peculiar characteristics in the soft X-ray domain, like a steep spectral continuum, 
and strong and fast variability. Most of these distinctive properties of NLS1s have been interpreted in the context of relatively small black hole masses 
undergoing rapid evolution, powered by higher accretion rates than their broad line equivalent \citep{peterson00}. 
The average hard X-ray photon index of the NLS1 in our sample is indeed slightly larger than for the rest of the Seyfert class 
($\langle \Gamma \rangle_{\rm NLS1} = 2.2 \pm 0.1$, $\langle \Gamma \rangle_{\rm Sey1} = 2.01 \pm 0.02$), as derived also from \textit{INTEGRAL} observations on 14 NLS1s 
\citep{ricci11,panessa11}.
On the other hand, the average variability amplitude is consistent with what is observed for broad line Seyfert 1 ($\langle S_{\rm V} \rangle_{\rm NLS1} = 15 \pm 3 \%$).
Even though our NLS1 sample is too small to draw any firm conclusion, the discrepancy between the soft and hard X-rays results could be due to the different time scales probed
in these studies: while NLS1s do show strong variability on short time scales, on long time scales above the PDS time break, their timing characteristics do not differ 
significantly from those of the general type-1 Seyfert population. 
This is in agreement with \citet{saxton11}, who found substantially the same long-term variability in narrow and broad line Seyfert 1 within their 0.2--2 keV survey study.

\subsection{Long-term variability, luminosity and black hole mass}

Due to the rather long time-scales probed by BAT, the anti-correlations of variability with luminosity \citep{barr86,green93,papadakis04} and with black hole mass 
\citep{papadakis95,zhou10,kelly11,kelly13,ponti12} often observed below 10~keV on short time scales are not detected in our hard X-ray study (Figs.~\ref{Fig:Sv-lum} 
and \ref{Fig:Sv_Mbh}).
This is in agreement with results obtained in the soft X-ray band when year time scales are investigated.
For example, \citet{saxton11} did not find any relation between long-term variability and luminosity when studying a 0.2--2~keV AGN sample, and argued 
that this is expected since the sampled time scales are substantially longer than the typical PDS break times of AGN. Moreover, \citet{markowitz04} analysed a sample of 27 AGN
with \textit{ASCA} and \textit{RXTE} data in the 2--12~keV energy range. Even though they do detect anti-correlations between variability and luminosity and 
variability and black hole mass on day time scales, the significance of these relations decreases when studying longer time scales (a month to years)
as well as the slope of the linear fit to the data approaches zero, indicating that on long time scales there is no anti-correlation observed.
\citet{zhang11} used \textit{RXTE}/ASM data, monitoring 27 AGN for about 14 years in the 1.5--12~keV band. The excess variance computed on the 300-day binned \lcs\ does 
not correlate with the black hole mass. The author argues that this points to AGN being in a high/soft state rather than in a low/hard state, based on an analogy with 
Galactic black holes (GBH) when solely their timing behavior is considered. 
In fact, PDS of GBHs in low/hard state are observed to present a second break at even lower time frequencies, and therefore if a similar behavior 
has to be expected from AGN, a mass dependence of the excess variance should be observed also on long time scales, which is instead not detected. 

Finally, considering that the BAT PDS on long time scales appear to have a slope consistent with $\alpha_{\rm PDS} \sim 1$ (\sm), \Sv computed in this time frequency range
is proportional to the PDS normalization. Therefore, the lack of correlation between \Sv and $M_{\rm BH}$ and $L_{\rm X}$ indicates that the PDS normalization does not seem 
to scale with black hole mass nor luminosity.
Moreover, the average variability for Seyfert galaxies is $\langle S_{\rm V} \rangle_{\rm Sey} = 19 \%$ with 68\% of the objects having \Sv values in the range $\sim 10-35\%$. 
The total variability range is smaller by a factor of $\sim 6$ with respect to the range measured on shorter time scales by \citet{ponti12} in the 2--10~keV energy band.
This suggests that the BAT \Sv, and therefore the normalization of the PDS on long time scales, could be very similar in all local Seyfert galaxies. 

\subsection{Spectral variability}

A significant correlation is detected between variability in the 14--24 and 35--100~keV energy ranges, with Seyfert galaxies being
on average more variable in the lower energy band (Fig.~\ref{Fig:Sv_band}).
The detected correlation suggests that the same process is likely to be at the origin of the variability in the two bands. 
Moreover, the presence of such large average variations, even in the 35--100~keV range, is incompatible with hard X-ray variability being 
entirely due to variations of the hydrogen column density along the line of sight. In fact, as an indication, a variation of 
$\Delta N_{\rm H} \sim 5 \times 10^{24} \, \rm cm^{-2}$ would imply only a 3\% variation of the 35--100~keV flux. 
Yet, a limited fraction of the additional variability observed at lower (14--24~keV) energies could be due to absorption variations.

Several studies have detected decreasing variability with increasing X-ray energies in the 0.2--12~keV band, either through excess variance or PDS amplitude 
(e.g., \citealt{nandra01,mchardy04,markowitz04,ponti07,ponti12}), and both on short ($\lesssim 1 \, \rm day$) and long time scales ($\gtrsim 30 \, \rm days$).
The same finding extends to hard X-ray energies, with the 14--24~keV band showing on average larger variations than the 35--100~keV one in Seyfert galaxies. 
A similar result has been obtained by \citet{chitnis09} when comparing the 1.5--12~keV \textit{RXTE}/ASM variability to that measured in the 14--195~keV band by BAT.
\citet{shimizu13} did not detect an energy dependence of the PDS parameters for the 3 Seyferts for which this investigation was possible, but the corresponding 
excess variance (computed using the PDS parameters) decreases through the 14--24, 24--50 and 50--150~keV bands they studied.
If for observations below 10~keV changes of the amount of the absorbing material or of its ionisation could play a role \citep{risaliti02c,ponti12}, 
in the BAT band such mechanisms are unlikely to be able to fully explain the observed properties.

   \begin{figure}
   \centering
   \includegraphics[width=6.8cm]{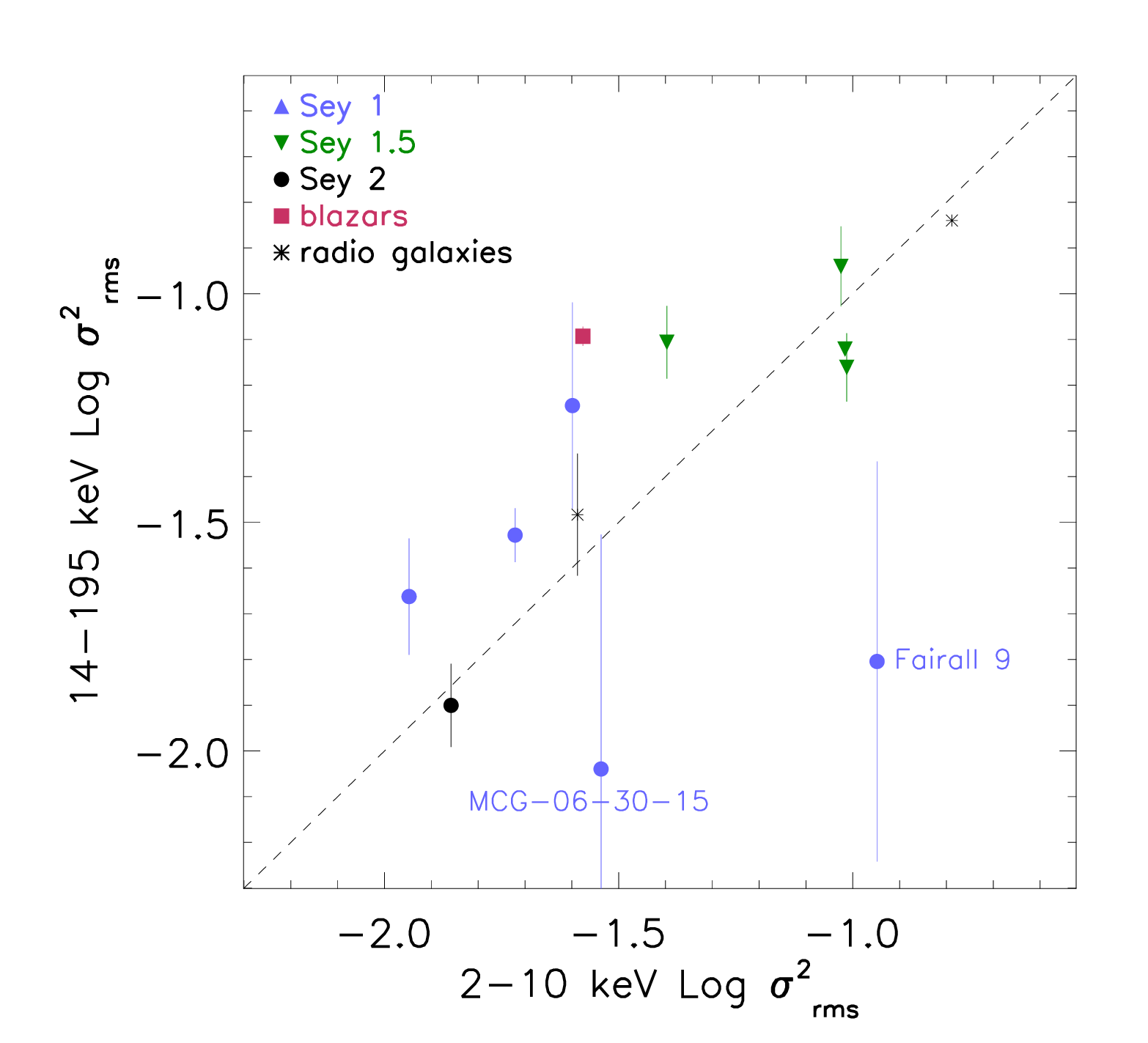}
      \caption{14--195~keV versus 2--10~keV normalized excess variance $\sigma^2_{\rm rms}$ for 10 Seyfert galaxies and 3 radio loud AGN (3C~273, 3C390.3, Cen~A) 
      in our sample for which 2--10~keV PDS with a measured frequency break have been reported in the literature. 
      The dashed line indicates where objects with the same variability at soft and hard X-rays would lie.
	      }
	 \label{Fig:soft}
   \end{figure}

The variations could instead originate directly at the X-ray source. For example, if a power law spectrum pivots at energies 
$\gtrsim 50$~keV, the 14--24~keV flux would be expected to vary more strongly than the 35--100~keV one \citep{gierlinski05}.
If we assume a standard Comptonization model based on a two-phase accretion disk \citep{haardt93}, the hard X-ray variability could be driven by changes 
of the seed photon flux \citep{nandra00,arevalo05} or by physical variations in the corona (temperature and optical depth; \citealt{nicastro00,lubinski10}).
\citet{petrucci00} point out that in order to have a pivoting point at high energies an increase of the
cooling is required rather than a decrease of the heating, that would instead determine a low-energy pivoting point.

In addition, the superposition of two components, one constant (e.g., the reflected continuum) and the other one variable
(for example the Comptonized continuum) could further contribute to the observed spectral variations \citep{shih02,markowitz04,ponti06,papadakis09,sobolewska09}.
In the 14--24~keV and 35--100~keV bands the contribution of reflection to the observed flux is about 30--40\% for a reflection fraction $R=1$ and typical AGN spectra with 
$\Gamma = 1.6-2.1$, $N_{\rm H} = 1-100 \times 10^{22} \rm \, cm^{-2}$, and a high-energy cut-off at $E_{\rm C} = 100-300 \rm \, keV$. 
For such spectra, a larger contribution from reflection is measured in the total 35--100~keV flux compared to the 14--24~keV one, but only by a few percents. 
Therefore, reflection alone cannot fully explain the difference observed in the two hard X-ray bands, even though a contribution from it is expected.

The scenario of a pivoting spectrum superposed to a constant reflection is in general also consistent with a steepening of the hard X-ray spectrum when the source brightens. 
This is indeed observed for the majority of the Seyferts in our sample that show some spectral variability (Fig.~\ref{Fig:HR_flux}) and follows an established trend observed in 
the X-ray emission of numerous local radio quiet AGN \citep{markowitz04,sobolewska09,caballero11} and also in higher-redshift objects \citep{paolillo04}. 
Within the Comptonization frame, the softer-when-brighter behavior can be understood if an increase of the seed photon power illuminating the corona 
determines a more efficient cooling of the hot electrons, with a resulting drop of the coronal temperature, which causes the X-ray spectrum to steepen.

The correlation between Eddington ratio and photon index is another known property of AGN samples below 10~keV \citep{sobolewska09} and it is confirmed by our study 
at hard X-rays (Fig.~\ref{Fig:Gamma_Sv}, middle panel), in agreement with \citet{middleton08} who suggested this relation to explain possible differences in 
hard X-ray spectral slopes for different AGN classes.

A different behavior arises when comparing the hard X-ray to the soft X-ray variability. We collected from the literature the PDS parameters for 13 AGN in our sample, 
i.e., those objects with measured soft X-ray PDS and with detected frequency break (\citealt{gonzalez12}, \sm, and references therein). 
We compare the normalized excess variance $\sigma^2_{\rm rms}$ in the 14--195~keV band to the 2--10~keV one, 
computed by extrapolating the soft X-ray PDS to low frequencies in order to cover the same time-scales as the BAT data set, i.e. between 
$\nu_{\rm min} = 6 \times 10^{-9} \rm \, Hz$ and $\nu_{\rm max} = 4 \times 10^{-7} \rm \, Hz$. The variability amplitude at hard X-rays is found to be equal or larger 
than that in the soft X-rays (Fig.~\ref{Fig:soft}). This is consistent with what has been found by \citet{caballero11} and \citet{shimizu13}, yet it is surprising, 
as it seems to be an opposite trend to those observed within the soft and hard X-rays bands alone.
One reason could be that, when extrapolating the 2--10~keV PDS to year time-scales, the uncertainties on the measurement of the PDS slope introduce
a large scatter in the estimated excess variance. 
Indeed, for 6 out of 13 objects the low-frequency PDS slope has been fixed to $\alpha=-1$ (from \citealt{gonzalez12}). 
Slightly steeper slopes by $\Delta \alpha = 0.1-0.2$ would imply a comparable or larger variability at soft rather than at hard X-rays.

If instead the hard X-rays are indeed more variable than the soft X-rays on long time scales, this might indicate an even more complex dependence of variability with energy.
The difference between soft and hard X-ray variability could be, for example, related to the size of the emitting region. 
If the high-energy plasma is located inside the accretion disk or above its inner part \citep{zdziarski99,lubinski10}, while the low-energy emitting region is associated to 
some outflow from the disk or a second plasma with lower temperature \citep{petrucci13}, then the high-energy emission could be more variable than the lower-energy emission.
However, the change in relative dominance of the two components would have to happen exactly at energies of 10--15~keV. 
In addition, it would be necessary to explain why within the soft and hard X-ray bands this trend is inverted, with the higher-energy emission being in general 
less variable than the lower-energy emission.

   \begin{figure}
   \centering
   \includegraphics[width=6.8cm,angle=90]{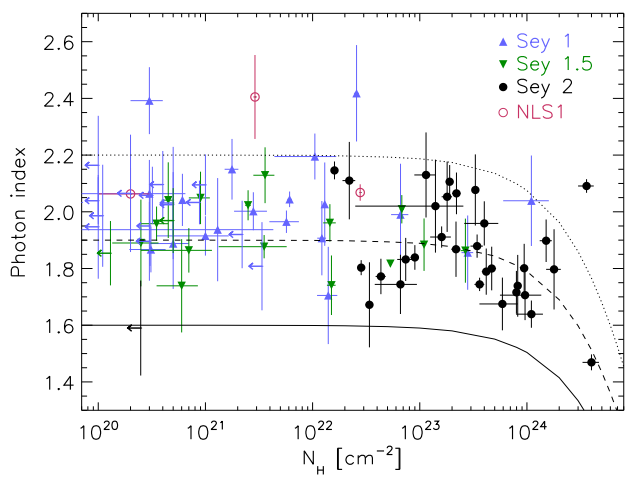}
      \caption{Measured photon index versus intrinsic absorption for the Seyfert galaxy sample. 
      		The lines correspond to the measured photon index obtained from a simple power law fit of simulated power law spectra with intrinsic photon index
      		$\Gamma$ = 1.6, 1.9 and 2.2 (from bottom to top) modified by the corresponding $N_{\rm H}$. Upper limits on $N_{\rm H}$ are indicated with arrows. 
	      }
	 \label{Fig:Gamma_Nh}
   \end{figure}

\subsubsection{Variability and photon index anti-correlation}

Another interesting difference compared to the lower energies and lower time frequencies is the trend of harder sources being more variable at hard X-rays 
(Fig.~\ref{Fig:Gamma_Sv}, right panel). In fact, several papers report the opposite trend of softer sources being more variable on time scales of the order or below a day, 
even though with different degrees of confidence. \citet{oneill05} found a marginal correlation between photon index and excess variance, while \citet{green93} and 
\citet{grupe01} found a more significant correlation in the 1--8~keV (\textit{EXOSAT}) and 0.2--2~keV band (\textit{ROSAT}), respectively. 
One has to keep in mind though that a contribution from absorption variations could be important in the \textit{ROSAT} data.
The largest soft X-ray sample for this kind of study has been presented by \citet{ponti12}. They confirm the correlation between excess variance and photon index, which 
they ascribe to the the fact that accretion rate correlates with both variability and photon index in their sample. 
On the contrary, at hard X-rays and low time frequency, it seems more likely that the correlation between accretion rate and photon index and the anti-correlation between 
photon index and variability are at the origin of the marginal anti-correlation between accretion rate and variability (Fig.~\ref{Fig:Gamma_Sv}).

The photon index versus variability anti-correlation is instead commonly observed in Galactic black holes (GBH), which show larger variability during the
hard state, i.e., when the corona rather than the accretion disk emission is believed to dominate the X-ray spectrum (e.g., \citealt{munozdarias11}).
Moreover, a decreasing variability with the softening of the spectrum is detected in the so-called hard-intermediate state, when the less variable accretion
disk component starts contributing to the X-ray spectrum of GBHs \citep{belloni11}. 
This might indicate that the larger variability in the harder BAT sources could be associated to their stronger Comptonization.

The anti-correlation between variability and photon index in our sample could be, at least partly, induced by the effect of different reflection fractions. 
On one hand, if one assumes that the bulk of reflection is mostly constant, objects with larger reflection would be expected to be less variable. 
Constant or slowly variable reflection components observed in a number of objects have been explained either due to a distant reflector, with light travel time damping
out variability (e.g., \citealt{miller08}, \citealt{bianchi09}), or as a consequence of strong gravitational light bending in the close vicinity of the black hole
(e.g., \citealt{miniutti04}, \citealt{ponti06}, \citealt{miniutti07}).
On the other hand, there are indications that objects with larger reflection have softer X-ray spectra. In fact, \citet{zdziarski99} reported a strong correlation indicating 
that intrinsically softer sources have larger reflection fractions, possibly due to the reflector being also the source of seed photons for the Comptonization emission.
In addition to this effect, when fitting with a simple power law model spectra with different reflection strengths and the same intrinsic continuum, the measured photon index 
in the 14--195 keV band results steeper for larger reflection, with $\Delta \Gamma \lesssim 7\%$ (depending on the intrinsic continuum shape and on the position of the 
high-energy spectral turnover) for a reflection fraction $R$ changing from 0 to 2.
Reflection fractions $R > 1$ might be the result of peculiar geometries (causing, e.g., high covering factors; \citealt{malzac01}), 
of light bending effects in the vicinity of the black hole \citep{miniutti04}, or of a variable continuum combined with a delayed answer of 
the reflected component, due to the distance of the reflector \citep{malzac02}.

When the accretion disk is considered as the main reflector, a larger reflection component is expected from face-on compared to edge-on AGN. Indeed, when dividing our sample 
into the different AGN classes, Seyfert~1 appear at the higher end of the photon index distribution, consistent with larger reflection being associated to softer spectra.
Furthermore, \citet{ricci11} and \citet{vasudevan13} have recently shown that objects with absorption in the range $10^{23} < N_{\rm H} < 10^{24} \rm \, cm^{-2}$ present a 
stronger reflection component than those at lower $N_{\rm H}$. When looking at the distribution of photon indexes as a function of $N_{\rm H}$, sources in the range 
$10^{23} < N_{\rm H} < 4 \times 10^{23} \rm \, cm^{-2}$ have a softer spectrum than the rest of the Seyfert~2 population and comparable to the average Seyfert 1 one 
(Fig.~\ref{Fig:Gamma_Nh}), possibly pointing out a stronger reflection component. 
We note that the photon index drop at $N_{\rm H} > 4 \times 10^{23} \rm \, cm^{-2}$ cannot be simply explained by the fact that our fitting does not take into account absorption. In fact, 
when simulating absorbed spectra and fitting them with a simple power law, the measured photon index decreases much less dramatically as a function of $N_{\rm H}$ then 
observed in the data (see lines in Fig.~\ref{Fig:Gamma_Nh}).

However, caution should be used considering that for highly absorbed sources simple absorption models fail to correctly reproduce the spectra in presence of large reprocessed 
emission (Lubinski et al. in preparation, \citealt{yaqoob12}, \citealt{braito13}), and that stacked spectra might be influenced by the effects of combining objects with 
different spectral shapes contributing with varying weight to the different energy bands.
In addition, the validity of the $\Gamma-R$ correlation reported by \citet{zdziarski99} is still debated. Some later studies confirmed this relation 
\citep{mattson07,dadina08,winter12} whereas others did not \citep{molina09,derosa11}, pointing out the difficulty in disentangling a real physical relation from the 
degeneracies between photon index, reflection and high-energy cut-off in the commonly used X-ray fitting models \citep{mattson07,winter12}.

An alternative explanation could come from a similar trend observed in NGC~7469 \citep{nandra01b}, which presented a higher excess variance when the spectrum of the source 
was harder. \citet{nandra01b} suggest that this and other timing properties of this AGN could be explained if the inner regions of the corona are hotter than the outer ones. 
The hard X-ray emission would then be produced in the innermost parts of the corona and therefore, be more variable and on shorter time scales than the soft X-ray one, 
produced in the larger and cooler outer zone. 
However, these results are based on a 30 days long observation at $<$15~keV, hence the time scales and energies are different than those of our work.
Similarly, in the model first proposed by \citet{lyubarskii97} and then reconsidered by others \citep{churazov01,arevalo06}, different variability patterns at different 
energies and on different time scales are explained with accretion disk fluctuations propagating inward and illuminating a corona with emissivity gradient, 
which would therefore modulate the X-ray source. \\


\section{Conclusions}\label{section:conclusions}

We have characterized the 14--195~keV variability of the largest sample of hard X-ray selected AGN provided by \textit{Swift}/BAT up to date.
The amplitude of the variations has been quantified and their dependence on energy and on basic AGN properties, such as black hole mass, 
luminosity, Eddington ratio and intrinsic absorption, has been investigated.
The average characteristics of hard X-ray variability are in general agreement with the unified models, indicating larger variability in jetted sources
and no clear dependence of the variations on AGN orientation.
The amplitude of the hard X-ray variations and its spectral properties indicate that variability cannot be driven by absorption variations. On the other hand
for radio quiet AGN it is rather consistent with changes of a power law continuum over a constant component, with the power law pivoting around $\gtrsim 50 \rm \, keV$.
Such variations of the continuum emission would be achieved for example with a change in the corona temperature, determining a shift of the high-energy cut-off at hard X-rays, 
either due to intrinsic coronal variations or to a change in the seed photon flux.

In general, variability at hard X-rays has very similar properties to that in the softer X-ray band, when similar time scales are compared, compatible
with variability being driven by the same mechanism across the full X-ray band.
However, some discrepancies, like AGN with harder spectra being more variable and the apparent lower variability in the 2--10~keV band, could be ascribed to additional
components in either band (e.g., reflection, absorption), and in any case could provide interesting insights on the intrinsic differences of the emission in the 
soft and hard X-ray bands, as for example concerning the emitting regions and their sizes. 

The lack of dependence of the long-term variability on black hole mass and luminosity, and the relatively narrow range of variability measured in Seyfert galaxies suggest that 
the normalization of the X-ray power spectrum is independent on black hole mass and luminosity and might have a similar value for all local Seyfert galaxies.

Thanks to the \textit{Swift}/BAT on-going observations it will be possible to further extend the studied time scales and increase the available AGN sample, 
obtaining the best hard X-ray AGN variability sample for some time to go. At the same time, pointed observations with \textit{Suzaku}, \textit{INTEGRAL} and \textit{NuSTAR}
will continue to investigate the spectral variability in bright and well studied AGN as already successfully done in a number of sources
\citep{reeves07b,itoh08,terashima09,fukazawa11,beckmann07c,soldi09,soldi11a,lubinski10}.

\begin{acknowledgements}
      The authors thank Jerome Rodriguez for useful discussions on Galactic black holes, 
      and the anonymous referee for the valuable suggestions that helped to improve this paper.
      SS acknowledges the Centre National d'Etudes Spatiales (CNES) for financial support.
      GP acknowledges support via an EU Marie Curie Intra-European fellowship under contract no. FP-PEOPLE-2012-IEF-331095.
      This work has been partly supported by the LabEx UnivEarthS\footnote{http://www.univearths.fr/en} project ``Impact of black holes on their environment'', 
      and by the Polish NCN grants N N203 581240 and 2012/04/M/ST9/00780.
      This article commemorates our colleague Jack Tueller who passed away during the study.
\end{acknowledgements}

\bibliographystyle{aa}
\bibliography{biblio}

%
\setlength{\LTcapwidth}{7in}
\onecolumn
\begin{table}
\caption{\label{table:average} Average properties of the BAT AGN in our variability sample, as a function of the AGN class and the amount of intrinsic absorption. 
}
\begin{tabular}{lcccccc}
\noalign{\smallskip}
\hline\hline
\noalign{\smallskip}
 Type & $\langle z \rangle^a$  & $\langle F_{14-195 \rm \, keV} \rangle^a$ & $\langle \Gamma \rangle^a$ & $ \langle S_{\rm V} \rangle^a$ & $ \langle S_{\rm V, 14-24 \, keV} \rangle^b$ & $ \langle S_{\rm V, 35-100 \, keV} \rangle^b$ \\
  &    & [$10^{-11} \rm \, erg \, cm^{-2} \, s^{-1}$] &  &      [\%]      &    [\%]     & [\%] \\
\hline
\noalign{\smallskip}
Sey 1 & 0.033 (36) &  6.4 $\pm$ 0.8 & 2.01 $\pm$0.02&           16.5 $\pm$ 0.9
 &           21 $\pm$ 2 (15) &           13 $\pm$ 2 \\
Sey 1.5 & 0.017 (17) & 10 $\pm$ 2 & 1.92 $\pm$0.03&   24 $\pm$ 1
&           32 $\pm$ 2 (11) &           23 $\pm$ 2 \\
Sey 2 & 0.016 (32) &10 $\pm$ 1 & 1.86 $\pm$0.03 &         	 19.9 $\pm$ 0.7
 &           28 $\pm$ 1 (22) &           16.2 $\pm$ 0.9 \\
NLS1 & 0.019 (3) &11 $\pm$ 6 &2.2 $\pm$0.1 &         	 15 $\pm$ 3
 &           24 $\pm$ 4 (2) &            9 $\pm$ 4 \\
\noalign{\smallskip}
\hline
\noalign{\smallskip}
unabs Sey & 0.031 (41) & 6.7 $\pm$ 0.7 &2.00 $\pm$0.02 &18.5 $\pm$ 0.8
 &           25 $\pm$ 2 (23) &           17 $\pm$ 2 \\
abs Sey & 0.016 (46) &10 $\pm$ 1 &1.90 $\pm$0.03 &          	   19.7 $\pm$
 0.6 &           28 $\pm$ 1 (26) &           16.4 $\pm$ 0.8 \\
CT Sey & 0.011 (6) &13 $\pm$ 4 &1.8 $\pm$0.1 &               		  16
 $\pm$ 1 &           37 $\pm$ 6 (4) &            9 $\pm$ 1 \\
\noalign{\smallskip}
\hline
\noalign{\smallskip}
all Sey & 0.023 (88) & 8.5 $\pm$ 0.8 &1.95 $\pm$0.02 &          	   19.3 $\pm$
 0.5 &           26.9 $\pm$ 0.9 (50) &           16.7 $\pm$ 0.7 \\
radio galaxies & 0.033 (9) &22 $\pm$ 12 &2.1 $\pm$ 0.2 &      	 24
 $\pm$ 1 &           26 $\pm$ 2 (9) &           26 $\pm$ 2 \\
blazars/quasars & 0.855 (13) & 7 $\pm$  3 &2.0 $\pm$ 0.2 &  33 $\pm$
 2  &           36 $\pm$ 2 (9) &           35 $\pm$ 2 \\
\hline\hline
\end{tabular}
\label{average}
\tablefoot{In parentheses, the number of objects used for the given average value is indicated. $^a$ All objects in the sample have measured redshift and photon index 
(from a simple power law fit).  $^b$ Only objects having variability measured in both the 14--24 and the 35--100~keV bands have been considered here.}\\
\end{table}
%
\setlength{\LTcapwidth}{7in}
\clearpage
\onecolumn
\begin{longtab}
\begin{longtable}{l c c c c c c c}
\caption{\label{table:Sv} Variability estimator $S_{\rm V}$, black hole mass and bolometric luminosity for the selected \textit{Swift}/BAT AGN sample.} \\
\hline\hline
\noalign{\smallskip}
Source name & Type & $\bar{x}$                             & $S_{\rm V}$ & $S_{\rm V, \, 14-24 \, keV}$ & $S_{\rm V, \, 35-100 \, keV}$ & $\log M_{\rm BH}$ &  $\log L_{\rm bol}$ \\ 
            &      & [$10^{-4}$ c/s/det] &      [\%]     &       [\%]    &     [\%]      & [$M_{\sun}$] & [$\rm erg \, s^{-1}$]  \\ 
\noalign{\smallskip}
\hline
\noalign{\smallskip}
\endfirsthead
\caption{continued.}\\
\hline\hline
\noalign{\smallskip}
Source name & Type & $\bar{x}$                             & $S_{\rm V}$ & $S_{\rm V, \, 14-24 \, keV}$ & $S_{\rm V, \, 35-100 \, keV}$ & $\log M_{\rm BH}$ &  $\log L_{\rm bol}$ \\ 
            &      & [$10^{-4}$ c/s/det] &      [\%]     &       [\%]    &     [\%]      & [$M_{\sun}$] & [$\rm erg \, s^{-1}$]  \\ 
\noalign{\smallskip}
\hline
\noalign{\smallskip}
\endhead
\hline
\endfoot
QSO B0033$+$595                   &  BLLac  &    0.60 &  $ 99 \pm  19 $  &  
$  94 \pm  16 $  &  $ 127 {+ 47 \atop - 52}$  &  \ldots  &  \ldots  \\
NGC 235A                        &  Sey1  &    0.74 &  $ 14 {+  6 \atop -  7}$
  &  $ \ldots$  &  $  10 {+  5 \atop -  6}$  &  $ 8.8 \pm 0.5^a$  &  \ldots  \\
Mrk 348                         &  Sey2  &    2.56 &  $ 25 \pm   3 $  &  
$  30 \pm   4 $  &  $  22 \pm   4 $  &  $ 7.2 \pm 0.7^b$  &  46.4$^f$  \\
Fairall 9                       &  Sey1  &    0.84 &  $ 10 {+  4 \atop -  5}$
  &  $ \ldots$  &  $ \ldots$  &  $ 8.9 \pm 0.5^a$  &  44.9$^o$  \\
NGC 526A                        &  Sey1.5  &    0.91 &  $ 35 \pm   6 $  &  
$  43 {+  9 \atop - 10}$  &  $  26 \pm  10 $  &  $ 8.0 \pm 0.5^a$  &  \ldots  \\
NGC 612                         &  NLRG  &    0.76 &  $ 27 {+  6 \atop -  7}$
  &  $  38 {+ 10 \atop - 12}$  &  $   9 \pm   5 $  &  $ 8.5 \pm 0.5^a$  &  
\ldots  \\
ESO 297$-$018                     &  Sey2  &    1.01 &  $ 16 \pm   4 $  &  
$  27 \pm   8 $  &  $ \ldots$  &  $ 9.7 \pm 0.5^b$  &  44.2$^j$  \\
NGC 788                         &  Sey2  &    1.19 &  $ 15 \pm   4 $  &  
$  19 {+  7 \atop -  8}$  &  $ \ldots$  &  $ 7.5 \pm 0.7^b$  &  44.3$^f$  \\
LEDA 138501                     &  Sey1  &    0.88 &  $ 24 \pm   7 $  &  
$  17 {+  8 \atop -  9}$  &  $  24 {+ 12 \atop - 13}$  &  \ldots  &  \ldots  \\
NGC 931                         &  Sey1.5  &    1.07 &  $ 12 \pm   5 $  &  
$ \ldots$  &  $ \ldots$  &  $ 7.6 \pm 0.3^b$  &  44.5$^f$  \\
$[$HB89$]$ 0241$+$622  &  Sey1  &    1.38 &  $ 15 {+  4 \atop -  5}$  &  
$  18 \pm   8 $  &  $  14 {+  6 \atop -  7}$  &  \ldots  &  \ldots  \\
NGC 1142                        &  Sey2  &    1.31 &  $ 30 \pm   4 $  &  
$  31 {+  8 \atop -  9}$  &  $  30 {+  6 \atop -  7}$  &  $ 9.4 \pm 0.5^b$  &  
44.8$^j$  \\
NGC 1275                        &  RG  &    1.65 &  $ 15 {+  4 \atop -  5}$  &  
$  14 \pm   4 $  &  $  25 {+ 13 \atop - 14}$  &  $ 8.5 \pm 0.7^b$  &  
45.0$^f$  \\
NGC 1365                        &  Sey1.8  &    1.09 &  $ 22 \pm   4 $  &  
$  26 {+  6 \atop -  7}$  &  $  10 \pm   5 $  &  $ 7.7 \pm 0.3^b$  &  
43.8$^j$  \\
ESO 548$-$G081                    &  Sey1  &    0.68 &  $ 36 {+  8 \atop -  9}$
  &  $  51 {+ 16 \atop - 15}$  &  $  33 \pm  14 $  &  $ 8.9 \pm 0.5^a$  &  
44.0$^j$  \\
3C 111.0                        &  RG  &    1.91 &  $ 16 {+  4 \atop -  5}$  &  
$  15 \pm   6 $  &  $  20 {+  7 \atop -  8}$  &  $ 9.6 \pm 0.8^b$  &  \ldots  \\
3C 120                          &  RG  &    1.56 &  $ 23 \pm   4 $  &  
$  23 {+  6 \atop -  7}$  &  $  21 {+  7 \atop -  8}$  &  $ 7.83 \pm 0.04^c$
  &  45.3$^f$  \\
UGC 03142                       &  Sey1  &    0.77 &  $ 34 {+  9 \atop - 10}$
  &  $  56 {+ 21 \atop - 22}$  &  $ \ldots$  &  \ldots  &  \ldots  \\
2MASX J04440903$+$2813003         &  Sey2  &    0.81 &  $ 91 \pm  17 $  &  
$ 104 {+ 20 \atop - 19}$  &  $  73 {+ 21 \atop - 20}$  &  \ldots  &  \ldots  \\
1RXS J045205.0$+$493248           &  Sey1  &    1.03 &  $ 21 {+  7 \atop -  8}$
  &  $  20 {+ 10 \atop - 11}$  &  $ \ldots$  &  $ 8.0 \pm 0.3^d$  &  
45.0$^d$  \\
2MASX J05054575$-$2351139         &  Sey2  &    0.87 &  $ 33 \pm   6 $  &  
$  28 {+  8 \atop -  9}$  &  $  16 \pm   7 $  &  $ 7.5 \pm 0.5^a$  &  \ldots  \\
IRAS 05078$+$1626                 &  Sey1.5  &    1.50 &  
$ 17 {+  6 \atop -  8}$  &  $   3 \pm   2 $  &  $ \ldots$  &  
$ 6.98 {+0.01 \atop -0.2}^e$  &  \ldots  \\
Ark 120                         &  Sey1  &    1.07 &  $  9 {+  4 \atop -  5}$
  &  $ \ldots$  &  $ \ldots$  &  $ 8.18 {+0.05 \atop -0.06}^b$  &  44.9$^f$  \\
ESO 362$-$18                      &  Sey1.5  &    0.76 &  $  9 \pm   4 $  &  
$  19 \pm   9 $  &  $ \ldots$  &  $ 9.0 \pm 0.5^a$  &  \ldots  \\
PICTOR A                        &  BLRG  &    0.63 &  $ 51 {+ 10 \atop -  9}$
  &  $  33 {+ 14 \atop - 13}$  &  $  69 {+ 18 \atop - 17}$  &  $ 7.6 \pm 0.5^a$
  &  \ldots  \\
PKS 0521$-$36                     &  BLLac  &    0.52 &  $ 29 {+  9 \atop - 10}$
  &  $  55 \pm  17 $  &  $ \ldots$  &  $ 8.7 \pm 0.3^f$  &  \ldots  \\
PKS 0548$-$322                    &  BLLac  &    0.55 &  $ 24 {+  9 \atop - 10}$
  &  $  15 {+  7 \atop -  8}$  &  $  30 {+ 15 \atop - 14}$  &  $ 8.2 \pm 0.3^f$
  &  \ldots  \\
NGC 2110                        &  Sey2  &    4.77 &  $ 33 \pm   3 $  &  
$  35 \pm   4 $  &  $  32 \pm   4 $  &  $ 8.3 \pm 0.3^b$  &  44.1$^f$  \\
MCG $+$08$-$11$-$011                  &  Sey1.5  &    2.15 &  
$ 33 {+  5 \atop -  4}$  &  $  32 \pm   5 $  &  $  31 \pm   6 $  &  
$ 8.1 \pm 0.6^b$  &  45.0$^j$  \\
2MASX J05580206$-$3820043         &  Sey1  &    0.60 &  $ 29 {+  9 \atop - 10}$
  &  $  35 {+ 11 \atop - 12}$  &  $ \ldots$  &  $ 8.4 \pm 0.5^a$  &  \ldots  \\
IRAS 05589$+$2828                 &  Sey1  &    1.14 &  $ 33 \pm   8 $  &  
$  25 \pm  10 $  &  $ \ldots$  &  \ldots  &  44.7$^d$  \\
Mrk 3                           &  Sey2  &    1.86 &  $ 35 {+ 11 \atop -  9}$
  &  $  54 {+ 20 \atop - 15}$  &  $  29 {+  6 \atop -  5}$  &  $ 8.7 \pm 0.3^b$
  &  44.5$^f$  \\
Mrk 6                           &  Sey1.5  &    0.93 &  $ 18 {+  5 \atop -  6}$
  &  $ \ldots$  &  $  12 {+  6 \atop -  7}$  &  $ 8.13 \pm 0.04^c$  &  
44.3$^j$  \\
Mrk 79                          &  Sey1.2  &    0.75 &  $  6 \pm   4 $  &  
$  29 {+  9 \atop - 10}$  &  $ \ldots$  &  $ 8.4 \pm 0.5^a$  &  44.6$^f$  \\
2MASS J07594181$-$3843560         &  Sey1.2  &    0.82 &  $ 10 \pm   5 $  &  
$  31 \pm  12 $  &  $  17 {+  8 \atop -  9}$  &  $ 8.3 \pm 0.5^b$  &  \ldots  \\
Mrk 1210                        &  Sey2  &    0.91 &  $ 24 {+  8 \atop -  9}$
  &  $   9 \pm   6 $  &  $  40 \pm  14 $  &  $ 6.5 \pm 0.7^g$  &  \ldots  \\
Fairall 272                     &  Sey2  &    0.64 &  $ 50 {+ 12 \atop - 13}$
  &  $  47 {+ 25 \atop - 26}$  &  $  36 {+ 12 \atop - 13}$  &  \ldots  &  
\ldots  \\
$[$HB89$]$ 0836$+$710  &  blazar  &    0.90 &  $ 12 \pm   5 $  &  
$  12 {+  5 \atop -  7}$  &  $  22 {+  7 \atop -  8}$  &  $ 9.4 \pm 0.3^h$  &  
\ldots  \\
MCG $-$01$-$24$-$012                  &  Sey2  &    0.70 &  
$ 19 {+  9 \atop - 10}$  &  $  26 {+ 12 \atop - 11}$  &  $ \ldots$  &  
$ 7.2 \pm 0.5^a$  &  \ldots  \\
MCG $+$04$-$22$-$042                  &  Sey1.2  &    0.63 &  
$ 27 {+  8 \atop - 11}$  &  $ \ldots$  &  $  37 {+ 12 \atop - 14}$  &  
$ 8.5 \pm 0.5^a$  &  44.5$^d$  \\
Mrk 110                         &  NLS1  &    0.94 &  $ 12 \pm   5 $  &  
$  24 \pm   9 $  &  $ \ldots$  &  $ 7.42 {+0.09 \atop -0.1}^b$  &  44.7$^f$  \\
MCG $-$05$-$23$-$016                  &  Sey2  &    3.44 &  $ 19 \pm   3 $  &  
$  23 \pm   4 $  &  $  16 \pm   5 $  &  $ 6.3 \pm 0.5^b$  &  \ldots  \\
NGC 3081                        &  Sey2  &    1.28 &  $ 19 {+  6 \atop -  7}$
  &  $  18 {+  9 \atop - 10}$  &  $  17 \pm   9 $  &  $ 7.4 \pm 0.3^b$  &  
\ldots  \\
NGC 3227                        &  Sey1.5  &    1.81 &  $ 27 {+  5 \atop -  6}$
  &  $  33 {+  7 \atop -  8}$  &  $  25 \pm   6 $  &  $ 6.9 \pm 0.1^i$  &  
43.9$^f$  \\
NGC 3281                        &  Sey2  &    1.37 &  $ 18 {+  4 \atop -  5}$
  &  $  21 {+  8 \atop -  9}$  &  $   3 \pm   2 $  &  $ 8.0 \pm 0.5^b$  &  
44.2$^j$  \\
Mrk 421                         &  BLLac  &    3.17 &  $ 94 {+ 10 \atop -  9}$
  &  $  88 {+  9 \atop -  8}$  &  $ 107 \pm  12 $  &  $ 8.3 \pm 0.3^b$  &  
\ldots  \\
NGC 3516                        &  Sey1.5  &    1.77 &  $ 26 \pm   4 $  &  
$  39 \pm   5 $  &  $  20 {+  4 \atop -  5}$  &  $ 7.50 {+0.04 \atop -0.06}^i$
  &  44.3$^f$  \\
NGC 3783                        &  Sey1  &    2.84 &  $ 15 {+  2 \atop -  3}$
  &  $  20 \pm   3 $  &  $   9 \pm   4 $  &  $ 7.47 {+0.07 \atop -0.09}^b$  &  
44.4$^f$  \\
UGC 06728                       &  Sey1.2  &    0.51 &  $ 21 {+  8 \atop -  9}$
  &  $  15 {+  8 \atop -  9}$  &  $ \ldots$  &  $ 6.4 \pm 0.3^d$  &  
44.3$^d$  \\
2MASX J11454045$-$1827149         &  Sey1  &    0.86 &  $ 24 {+  8 \atop -  9}$
  &  $  25 \pm  13 $  &  $ \ldots$  &  $ 6.7 \pm 0.5^a$  &  \ldots  \\
LEDA 38038                      &  Sey2  &    0.85 &  $ 24 {+  8 \atop -  9}$
  &  $  30 {+ 12 \atop - 13}$  &  $ \ldots$  &  \ldots  &  \ldots  \\
NGC 4151                        &  Sey1.5  &    8.06 &  $ 29 {+  3 \atop -  4}$
  &  $  30 \pm   4 $  &  $  28 \pm   3 $  &  $ 7.5 {+0.1 \atop -0.6}^b$  &  
43.7$^f$  \\
NGC 4388                        &  Sey2  &    4.01 &  $ 26 \pm   2 $  &  
$  26 {+  2 \atop -  3}$  &  $  25 \pm   3 $  &  $ 7.2 \pm 0.6^b$  &  
43.7$^j$  \\
3C 273                          &  quasar  &    6.23 &  $ 29 \pm   3 $  &  
$  29 \pm   3 $  &  $  28 \pm   3 $  &  $ 9.81 {+0.1 \atop -0.07}^b$  &  
47.1$^o$  \\
NGC 4507                        &  Sey2  &    2.85 &  $ 20 \pm   4 $  &  
$  30 \pm   5 $  &  $  15 {+  3 \atop -  4}$  &  $ 7.6 \pm 0.6^b$  &  
44.3$^j$  \\
ESO 506$-$G027                    &  Sey2  &    1.28 &  $ 43 \pm   7 $  &  
$  43 \pm  10 $  &  $  31 \pm   7 $  &  $ 8.6 \pm 0.5^b$  &  44.6$^j$  \\
NGC 4593                        &  Sey1  &    1.40 &  $ 15 \pm   4 $  &  
$  17 {+  6 \atop -  7}$  &  $ \ldots$  &  $ 6.99 {+0.08 \atop -0.1}^b$  &  
44.1$^f$  \\
WKK 1263                        &  Sey1.5  &    0.72 &  $ 36 {+  8 \atop -  9}$
  &  $  54 \pm  19 $  &  $  40 {+ 14 \atop - 15}$  &  $ 8.0 \pm 0.5^b$  &  
44.3$^d$  \\
SBS 1301$+$540                    &  Sey1  &    0.52 &  $ 36 {+ 10 \atop -  9}$
  &  $ \ldots$  &  $  19 {+  9 \atop - 10}$  &  $ 7.5 \pm 0.5^a$  &  
44.9$^d$  \\
NGC 4945                        &  Sey2  &    3.48 &  $ 34 \pm   4 $  &  
$  48 \pm  10 $  &  $  28 \pm   4 $  &  $ 6.2 \pm 0.3^b$  &  \ldots  \\
ESO 323$-$077                     &  Sey1.2  &    0.73 &  
$ 19 {+  8 \atop -  9}$  &  $  42 {+ 20 \atop - 21}$  &  $ \ldots$  &  
$ 7.4 \pm 0.6^b$  &  \ldots  \\
Cen A                           &  RG  &   20.40 &  $ 39 \pm   4 $  &  
$  37 \pm   4 $  &  $  39 \pm   4 $  &  $ 8.0 \pm 0.6^b$  &  \ldots  \\
MCG $-$06$-$30$-$015                  &  Sey1.2  &    1.26 &  
$ 13 {+  6 \atop -  7}$  &  $  21 {+  8 \atop -  9}$  &  $ \ldots$  &  
$ 6.7 {+0.1 \atop -0.2}^b$  &  44.1$^d$  \\
NGC 5252                        &  Sey2  &    1.65 &  $ 59 {+  8 \atop -  7}$
  &  $  61 {+ 10 \atop -  9}$  &  $  54 \pm   8 $  &  
$ 9.03 {+0.4 \atop -0.02}^b$  &  45.4$^f$  \\
4U 1344$-$60                      &  Sey1.5  &    1.67 &  $ 14 \pm   4 $  &  
$  23 \pm   7 $  &  $   7 {+  4 \atop -  5}$  &  $ 7.4 \pm 0.1^j$  &  
44.1$^j$  \\
IC 4329A                        &  Sey1.2  &    4.82 &  $ 17 \pm   3 $  &  
$  31 \pm   5 $  &  $   9 {+  3 \atop -  4}$  &  $ 8.1 \pm 0.2^k$  &  
44.8$^f$  \\
Mrk 279                         &  Sey1.5  &    0.63 &  $ 36 {+  6 \atop -  7}$
  &  $  37 {+ 12 \atop - 13}$  &  $  35 {+ 11 \atop - 12}$  &  $ 8.6 \pm 0.5^a$
  &  45.0$^o$  \\
Circinus Galaxy                 &  Sey2  &    4.54 &  $ 11 \pm   1 $  &  
$  17 \pm   2 $  &  $ \ldots$  &  $ 6.3 \pm 0.1^l$  &  \ldots  \\
NGC 5506                        &  NLS1  &    4.12 &  $ 15 {+  3 \atop -  4}$
  &  $  25 {+  4 \atop -  5}$  &  $   8 \pm   4 $  &  $ 6.7 \pm 0.7^b$  &  
44.1$^j$  \\
NGC 5548                        &  Sey1.5  &    1.19 &  $ 31 {+  5 \atop -  4}$
  &  $  19 {+  7 \atop -  8}$  &  $  24 {+  6 \atop -  7}$  &  
$ 7.82 \pm 0.02^b$  &  44.8$^f$  \\
ESO 511$-$G030                    &  Sey1  &    0.76 &  $ 33 \pm  11 $  &  
$  53 {+ 15 \atop - 16}$  &  $ \ldots$  &  $ 8.7 \pm 0.5^b$  &  44.4$^j$  \\
NGC 5728                        &  Sey2  &    1.35 &  $ 27 {+  6 \atop -  7}$
  &  $  39 {+ 15 \atop - 24}$  &  $  19 {+  8 \atop -  9}$  &  $ 8.5 \pm 0.5^a$
  &  43.7$^j$  \\
Mrk 841                         &  Sey1  &    0.64 &  $ 21 \pm  10 $  &  
$  25 {+ 12 \atop - 13}$  &  $ \ldots$  &  $ 8.5 \pm 0.7^b$  &  45.8$^f$  \\
PKS 1510$-$08                     &  quasar  &    0.83 &  
$ 41 {+ 10 \atop - 12}$  &  $  89 {+ 27 \atop - 36}$  &  
$  27 {+ 11 \atop - 12}$  &  $ 8.7 \pm 0.5^f$  &  46.4$^f$  \\
VII Zw 653                      &  Sey1.2  &    0.87 &  $ 21 {+  7 \atop -  8}$
  &  $  32 {+ 10 \atop - 11}$  &  $  24 {+ 15 \atop - 16}$  &  \ldots  &  
\ldots  \\
Mrk 1498                        &  Sey1.9  &    0.74 &  $  5 \pm   3 $  &  
$ \ldots$  &  $ \ldots$  &  $ 8.6 \pm 0.5^a$  &  45.3$^j$  \\
2MASX J16481523$-$3035037         &  Sey1  &    0.77 &  $ 47 \pm   8 $  &  
$  54 {+ 16 \atop - 20}$  &  $  42 {+ 15 \atop - 16}$  &  \ldots  &  \ldots  \\
NGC 6240                        &  Sey2  &    1.06 &  $ 22 {+  6 \atop -  7}$
  &  $  54 {+ 17 \atop - 18}$  &  $   4 \pm   3 $  &  \ldots  &  \ldots  \\
Mrk 501                         &  BLLac  &    0.89 &  $ 61 \pm   8 $  &  
$  64 \pm   9 $  &  $  62 {+ 12 \atop - 13}$  &  $ 9.2 \pm 0.3^b$  &  \ldots  \\
1RXS J165605.6$-$520345           &  Sey1.2  &    0.69 &  $ 21 \pm  10 $  &  
$  26 {+ 15 \atop - 16}$  &  $  21 {+ 10 \atop - 12}$  &  $ 7.9 \pm 0.5^b$  &  
\ldots  \\
2MASS J16561677$-$3302127         &  blazar  &    0.87 &  $ 41 \pm   7 $  &  
$  44 {+ 19 \atop - 22}$  &  $  35 {+  8 \atop -  9}$  &  \ldots  &  \ldots  \\
NGC 6300                        &  Sey2  &    1.65 &  $ 17 \pm   3 $  &  
$  18 {+  5 \atop -  6}$  &  $   8 {+  4 \atop -  5}$  &  $ 5.5 \pm 0.4^b$  &  
42.9$^j$  \\
AX J1737.4$-$2907                 &  Sey1  &    1.96 &  $ 15 {+  3 \atop -  4}$
  &  $  33 \pm   7 $  &  $ \ldots$  &  $ 8.9 \pm 0.7^b$  &  \ldots  \\
IC 4709                         &  Sey2  &    0.66 &  $ 22 {+ 10 \atop - 11}$
  &  $  34 {+ 18 \atop - 19}$  &  $ \ldots$  &  \ldots  &  \ldots  \\
PKS 1830$-$21                     &  FSRQ  &    1.04 &  $ 18 {+  6 \atop -  8}$
  &  $  38 {+ 19 \atop - 21}$  &  $ \ldots$  &  \ldots  &  \ldots  \\
3C 382                          &  Sey1  &    1.47 &  $ 11 {+  3 \atop -  4}$
  &  $  15 {+  5 \atop -  6}$  &  $  11 \pm   5 $  &  $ 9.2 \pm 0.5^b$  &  
45.6$^o$  \\
ESO 103$-$035                     &  Sey2  &    1.88 &  $ 12 \pm   3 $  &  
$  17 \pm   5 $  &  $ \ldots$  &  $ 7.1 \pm 0.6^b$  &  44.6$^j$  \\
3C 390.3                        &  BLRG  &    1.69 &  $ 17 \pm   3 $  &  
$  16 \pm   5 $  &  $  12 \pm   5 $  &  $ 8.46 {+0.09 \atop -0.1}^b$  &  
44.9$^f$  \\
Fairall 51                      &  Sey1  &    0.70 &  $ 20 {+  7 \atop -  8}$
  &  $  45 {+ 17 \atop - 19}$  &  $ \ldots$  &  $ 7.5 \pm 0.5^m$  &  \ldots  \\
NGC 6814                        &  Sey1.5  &    1.19 &  $ 28 {+  7 \atop -  8}$
  &  $  36 {+ 14 \atop - 17}$  &  $  19 {+  8 \atop -  9}$  &  $ 7.1 \pm 0.2^b$
  &  43.9$^f$  \\
Cygnus A                        &  RG  &    2.13 &  $ 31 {+  4 \atop -  5}$  &  
$  32 {+  5 \atop -  6}$  &  $  35 \pm   6 $  &  $ 9.4 \pm 0.1^b$  &  
45.7$^j$  \\
QSO B1959$+$650                   &  BLLac  &    0.69 &  $ 35 {+  9 \atop - 10}$
  &  $  42 \pm   9 $  &  $ \ldots$  &  $ 8.1 \pm 0.3^b$  &  \ldots  \\
MCG $+$04$-$48$-$002                  &  Sey2  &    1.13 &  
$ 20 {+  6 \atop -  7}$  &  $  24 {+  8 \atop -  9}$  &  $ \ldots$  &  \ldots
  &  \ldots  \\
4C $+$74.26                       &  Sey1  &    0.87 &  $ 21 {+  6 \atop -  7}$
  &  $  14 {+  6 \atop -  8}$  &  $  28 \pm  11 $  &  $ 9.6 \pm 0.5^b$  &  
46.2$^f$  \\
Mrk 509                         &  Sey1.2  &    1.60 &  $ 17 {+  7 \atop -  8}$
  &  $  16 {+  8 \atop -  9}$  &  $  12 {+  6 \atop -  7}$  &  
$ 8.16 {+0.03 \atop -0.04}^b$  &  45.0$^f$  \\
IC 5063                         &  Sey2  &    1.07 &  $ 18 \pm   5 $  &  
$  38 \pm  13 $  &  $ \ldots$  &  $ 7.7 \pm 0.5^a$  &  44.5$^f$  \\
2MASX J21140128$+$8204483         &  Sey1  &    0.55 &  $ 12 \pm   6 $  &  
$  36 {+ 15 \atop - 16}$  &  $ \ldots$  &  $ 8.8 \pm 0.5^b$  &  45.5$^j$  \\
28P 206                         &  RG  &    3.20 &  $ 30 \pm   4 $  &  
$  29 {+  3 \atop -  4}$  &  $  31 \pm   5 $  &  \ldots  &  \ldots  \\
SWIFT J212745.6$+$565636          &  NLS1  &    0.73 &  $ 20 {+  7 \atop -  8}$
  &  $  21 \pm   9 $  &  $  20 \pm  11 $  &  $ 7.2 \pm 0.5^n$  &  \ldots  \\
6dF J2132022$-$334254             &  Sey1  &    0.72 &  $ 40 \pm   7 $  &  
$  49 {+ 12 \atop - 13}$  &  $  36 {+ 15 \atop - 16}$  &  \ldots  &  \ldots  \\
PKS 2149$-$306                    &  FSRQ  &    1.07 &  $ 26 \pm   7 $  &  
$ \ldots$  &  $  29 \pm  10 $  &  \ldots  &  \ldots  \\
NGC 7172                        &  Sey2  &    2.63 &  $ 28 \pm   3 $  &  
$  33 \pm   7 $  &  $  29 \pm   5 $  &  $ 7.7 \pm 0.6^b$  &  43.8$^j$  \\
NGC 7213                        &  Sey1.5  &    0.68 &  $ 23 {+  9 \atop - 10}$
  &  $ \ldots$  &  $ \ldots$  &  $ 8.6 \pm 0.5^a$  &  44.3$^f$  \\
NGC 7314                        &  Sey1  &    0.80 &  $ 21 {+  9 \atop - 10}$
  &  $ \ldots$  &  $ \ldots$  &  $ 6.0 \pm 0.5^b$  &  43.0$^j$  \\
3C 454.3                        &  FSRQ  &    1.74 &  $ 59 \pm   8 $  &  
$  60 {+ 11 \atop - 12}$  &  $  56 \pm   9 $  &  $ 9.2 \pm 0.7^b$  &  
47.3$^f$  \\
MR 2251$-$178                     &  Sey1  &    1.67 &  $ 10 {+  4 \atop -  5}$
  &  $  11 {+  5 \atop -  6}$  &  $   1 \pm   1 $  &  $< 6.9^b$  &  45.8$^o$  \\
NGC 7469                        &  Sey1  &    1.10 &  $ 23 \pm   6 $  &  
$  22 {+  9 \atop - 10}$  &  $  20 {+ 10 \atop - 11}$  &  $ 7.09 \pm 0.05^b$
  &  45.3$^f$  \\
Mrk 926                         &  Sey1.5  &    1.81 &  $ 27 {+  4 \atop -  5}$
  &  $  26 \pm   7 $  &  $  21 {+  5 \atop -  6}$  &  $ 7.1 \pm 0.6^b$  &  
\ldots  \\
NGC 7582                        &  Sey2  &    1.21 &  $ 23 \pm   5 $  &  
$  24 {+  7 \atop -  8}$  &  $  12 \pm   6 $  &  $ 8.3 \pm 0.5^a$  &  
43.3$^j$  \\
NGC 7603                        &  Sey1.5  &    0.82 &  $ 11 {+  6 \atop -  7}$
  &  $ \ldots$  &  $ \ldots$  &  $ 8.1 \pm 0.3^b$  &  44.7$^f$  \\
\end{longtable}
\tablefoot{The variability estimator is computed in the 14--195~keV ($S_{\rm V}$), 14--24~keV ($S_{\rm V, \, 14-24 \, keV}$), and 35--100~keV ($S_{\rm V, \, 35-100 \, keV}$) 
bands. $\bar{x}$ is the average count rate in the full 14--195~keV band. The bolometric luminosities have been estimated from the fitting of the spectral energy 
distribution (see Sect.~\ref{sect:bl_Nh_Edd}).
$^a$ \citet{winter09}; 
$^b$ \citet{beckmann09} and references therein;
$^c$ \citet{grier12};
$^d$ \citet{vasudevan09b};
$^e$ \citet{stalin11};
$^f$ \citet{woo02};
$^g$ \citet{bian07};
$^h$ \citet{kaspi07};
$^i$ \citet{denney10};
$^j$ \citet{vasudevan10};
$^k$ \citet{nikolajuk04};
$^l$ \citet{greenhill03};
$^m$ \citet{padovani88};
$^n$ \citet{malizia08};
$^o$ \citet{vasudevan07}.
}
\end{longtab}
\end{document}